\documentclass[aps,twocolumn,prd,showpacs,showkeys,preprintnumbers,superscriptaddress,nobibnotes,floatfix,longbibliography,notitlepage,nofootinbib]{revtex4-2}

\pdfoutput=1
\usepackage{amsmath}
\usepackage{amsfonts}
\usepackage{amssymb}
\usepackage{mathrsfs}
\usepackage{color}
\usepackage{slashed}
\usepackage{graphicx}
\usepackage{xcolor}
\usepackage{hyperref}
\hypersetup{colorlinks=true,allcolors=purple}
\usepackage{multirow, makecell}
\usepackage{listings}
\usepackage{upgreek}
\usepackage{bm}
\usepackage[capitalise]{cleveref}
\usepackage[normalem]{ulem}
\usepackage{siunitx}
\usepackage{booktabs}
\usepackage{tabularx}
\usepackage{physics}

\DeclareSIUnit \s {\second}
\DeclareSIUnit \ns {\nano\second}
\DeclareSIUnit \mus {\micro\second}
\DeclareSIUnit \ms {\milli\second}
\DeclareSIUnit \MB {\mega\byte}
\DeclareSIUnit \GB {\giga\byte}
\DeclareSIUnit \TB {\tera\byte}
\DeclareSIUnit \PB {\peta\byte}
\DeclareSIUnit \Mbps {\mega\bit/\s}
\DeclareSIUnit \Gbps {\giga\bit/\s}
\DeclareSIUnit \Tbps {\tera\bit/\s}
\DeclareSIUnit \Pbps {\peta\bit/\s}
\DeclareSIUnit \kton {\kilo\tonne} 
\DeclareSIUnit \kt {\kilo\tonne}
\DeclareSIUnit \Mt {\mega\tonne}
\DeclareSIUnit \eV {\electronvolt}
\DeclareSIUnit \keV {\kilo\electronvolt}
\DeclareSIUnit \MeV {\mega\electronvolt}
\DeclareSIUnit \GeV {\giga\electronvolt}
\DeclareSIUnit \PeV {\peta\electronvolt}
\DeclareSIUnit \EeV {\exa\electronvolt}
\DeclareSIUnit \ZeV {\zetta\electronvolt}
\DeclareSIUnit \m {\meter}
\DeclareSIUnit \cm {\centi\meter}
\DeclareSIUnit \in {\inchcommand}
\DeclareSIUnit \km {\kilo\meter}
\DeclareSIUnit \kV {\kilo\volt}
\DeclareSIUnit \kW {\kilo\watt}
\DeclareSIUnit \MW {\mega\watt}
\DeclareSIUnit \MHz {\mega\hertz}
\DeclareSIUnit \mrad {\milli\radian}
\DeclareSIUnit \year {years}
\DeclareSIUnit \POT {POT}
\DeclareSIUnit \sig {$\sigma$}
\DeclareSIUnit\parsec{pc}
\DeclareSIUnit\lightyear{ly}
\DeclareSIUnit\foot{ft}
\DeclareSIUnit\ft{ft}
\DeclareSIUnit \ppb{ppb}
\DeclareSIUnit \ppt{ppt}
\DeclareSIUnit \samples{S}
\DeclareSIUnit \pe{PE}
\DeclareSIUnit \mwe{mwe}

\newcommand{\enu}{\E_\enu}

\definecolor{MH}{rgb}{0.0,0.6,9}
\definecolor{NK}{rgb}{1,0.6,0.5}
\definecolor{palatinate}{rgb}{0.494, 0.192, 0.482}

\newcommand\percentage[2][round-precision = 0]{
    \SI[round-mode = places,
        scientific-notation = fixed, fixed-exponent = 0,
        output-decimal-marker={.}, #1]{#2e2}{\percent}%
}

\newcommand{\xUpperNone}{0.49}
\newcommand{\xUpperHalf}{0.69}
\newcommand{\xUpperFull}{1.01}
\newcommand{\dchisqnomNone}{13.54}
\newcommand{\dchisqnomHalf}{8.06}
\newcommand{\dchisqnomFull}{3.82}
\newcommand{\dchisqbestNone}{24.63}
\newcommand{\dchisqbestHalf}{15.74}
\newcommand{\dchisqbestFull}{8.52}
\newcommand{\dchisqworstNone}{7.65}
\newcommand{\dchisqworstHalf}{4.57}
\newcommand{\dchisqworstFull}{2.25}
\newcommand{\fLowerNom}{0.55}
\newcommand{\fLowerBest}{0.72}
\newcommand{\fLowerWorst}{0.24}

\newcommand{\f}{f_{\overline{\nu}/\nu}}

\begin{document}

\title{Implications of MicroBooNE's low sensitivity to electron antineutrino interactions in the search for the MiniBooNE excess}
\author{N.~W.~Kamp}
\affiliation{Massachusetts Institute of Technology; Cambridge, MA 02139, USA}
\author{M.~Hostert}
\affiliation{Perimeter Institute for Theoretical Physics, Waterloo, ON N2J 2W9, Canada}
\affiliation{School of Physics and Astronomy, University of Minnesota, Minneapolis, MN 55455, USA}
\affiliation{William I. Fine Theoretical Physics Institute, School of Physics and Astronomy, University of Minnesota, Minneapolis, MN 55455, USA}
\author{C.~A.~Arg\"uelles}
\affiliation{Department of Physics \& Laboratory for Particle Physics and Cosmology, Harvard University, Cambridge, MA 02138, USA}
\author{J.~M.~Conrad}
\affiliation{Massachusetts Institute of Technology; Cambridge, MA 02139, USA}
\author{M.~H.~Shaevitz}
\affiliation{Columbia University; New York, NY 10027, USA}

\date{\today}

\begin{abstract}
The MicroBooNE experiment searched for an excess of electron-neutrinos in the Booster Neutrino Beam (BNB), providing direct constraints on $\nu_e$-interpretations of the MiniBooNE low-energy excess (LEE).
In this article, we show that if the MiniBooNE LEE is caused instead by an excess of $\overline{\nu}_e$, then liquid argon detectors, such as MicroBooNE, SBND and ICARUS, would have poor sensitivity to it.
This is due to a strong suppression of $\overline{\nu}_e -{}^{40}$Ar cross sections in the low-energy region of the excess.
The MicroBooNE results are consistent at the $2\sigma$~C.L with a scenario in which the MiniBooNE excess is sourced entirely by $\overline{\nu}_e$ interactions.
The opportune location of ANNIE, a Gd-loaded water Cherenkov detector, allows for a direct search for a $\overline{\nu}_e$ flux excess in the BNB using inverse-beta-decay events.
\end{abstract}

\maketitle

\section{Introduction} \label{sec:intro}

The MiniBooNE Experiment at Fermi National Accelerator Laboratory (Fermilab) used a 450~t fiducial volume mineral-oil-based (CH$_2$) Cherenkov detector to search for the appearance of electron-like events in a beam made predominantly of muon-flavor neutrinos.
The beam, produced in the Booster Neutrino Beamline (BNB), resulted from 8.9~GeV total energy protons impinging on a beryllium target, with charged mesons magnetically focused toward the detector~\cite{MiniBooNE:2008hfu}.
The polarity of the magnet could be switched to allow either positively or negatively charged mesons to be focused.
The pions and kaons decayed to mainly produce a muon-flavor flux of neutrinos or antineutrinos, with low electron-flavor content, as discussed below.
The beam traversed largely undisturbed to reach the MiniBooNE detector located 541~m downstream.
The analysis sought to isolate the Charged-Current Quasielastic (CCQE) scattering of neutrinos, $\nu_e +n \rightarrow e^- + p$, or antineutrinos, $\overline \nu_e +p \rightarrow e^+ + n$.
In the MiniBooNE Cherenkov detector, both reactions appear as single electromagnetic-like Cherenkov rings.

During a series of runs from 2002 to 2019, the MiniBooNE experiment received $18.75\times10^{20}$ ($11.27 \times 10^{20}$) protons on target (POT) with the magnetic horn focusing positively (negatively) charged mesons.
An excess of $560.6 \pm 119.6$ ($77.4 \pm 28.5$) electron-like events above the background from intrinsic electron-flavor flux and misidentified muon-flavor events was observed~\cite{MiniBooNE:2020pnu}.
These low-energy excesses are often referred to as the MiniBooNE ``LEE'' signal.
The community has engaged in a thorough search for misidentified particles that were not included in the analysis but has not identified a conclusive explanation behind the full excess~\cite{Brdar:2021ysi,Kelly:2022uaa,Acero:2022wqg}.
This leads us to reconsider the flux that could cause the LEE signature.

Because the detector is limited to identifying an electron-like Cherenkov ring, it is not possible to identify the neutrino versus antineutrino content of the LEE.
Therefore, under the assumption that the LEE is caused by an excess of $\nu_e$ or $\overline{\nu}_e$ CCQE events in the detector, we can categorize the possible explanations as follows (defining $\f$ to be the $\overline{\nu}_e$ fractional contribution to the LEE in neutrino mode):
\begin{enumerate}
    \item \emph{Scenario 1}: The excess is entirely due to $\nu_e$ interactions in neutrino mode running, $\f = 0$, and entirely $\overline \nu_e$ in antineutrino mode running.
    This fits the classic model of sterile-enhanced $\nu_\mu \to \nu_e$ neutrino oscillations.
    \item \emph{Scenario 2}:
    The excess arises from a flux of mixed content.
    In this scenario, we assume the event rate of the excess is evenly split into neutrino and antineutrino events, $\f = 0.5$.
    In neutrino mode, this hypothesis corresponds to a flux excess of antineutrinos that is larger and lower-energy than the flux excess of neutrinos.
    \item \emph{Scenario 3}: The excess is entirely due to $\overline{\nu}_e$ interactions in neutrino and antineutrino mode running, $\f = 1$.
\end{enumerate}
Of these three possibilities, only \emph{Scenario 1} has been thoroughly explored by the community~\cite{Acero:2022wqg}.
While the two experiments were situated in very different beams, we note that an anomalous flux of antineutrinos in MiniBooNE may also be compatible with the unexplained signal at the LSND experiment.
LSND operated with a liquid scintillator detector at the LANSCE spallation source and observed a $3.8\sigma$-significant excess of inverse beta decay (IBD)~\cite{LSND:2001aii}.
Because of the unique IBD signature, a positron accompanied by delayed neutron capture, the LSND excess favors a $\overline \nu_e$ interpretation over a $\nu_e$ one.

Motivated by \emph{Scenario 1}, the MicroBooNE Experiment was proposed to run on the same Booster Neutrino Beamline (BNB), 70~m upstream of the MiniBooNE detector, using an 80~t fiducial volume liquid argon time projection chamber (LArTPC).
The LArTPC technology was selected in 2006 to greatly reduce photon-electron misidentification backgrounds that were thought, at the time, to be the best explanation of the MiniBooNE anomaly~\cite{MicroBooNE:2007ivj}.
However, as a state-of-the-art detector, the cost-per-ton for the detector led to a restricted size, and hence low statistics.
The experiment has published data taken from 2016-2018 totaling $6.9\times 10^{20}$~POT in neutrino mode.
In principle, the $\mathcal{O}(1)$~cm vertex resolution of the MicroBooNE detector~\cite{MicroBooNE:2020sar} and its ability to detect protons allows for the separation of  $\nu_e + n \rightarrow e^- + p$ and $\overline \nu_e +p \rightarrow e^+ + n$ events, making it ideal for testing mixed models like scenarios 2 and 3.
Unfortunately, in practice, the argon target has a highly suppressed antineutrino interaction cross section in the $E_{\overline{\nu}}< 600$~MeV range of interest for the LEE, as we explain below.
Thus, given the size of the detector and the length of the run, MicroBooNE is much less sensitive to an antineutrino component in the LEE.

MicroBooNE performed a model-agnostic search for the MiniBooNE LEE~\cite{MicroBooNE:2021rmx}, building a template of the excess with respect to the Standard Model prediction from the MiniBooNE neutrino-mode data.
The LEE template was derived by unfolding the difference between the central value of the data and background predictions at MiniBooNE to an excess of neutrinos in the beam.
This process assumed that the MiniBooNE LEE signal was entirely due to neutrino interactions.
The strategy to obtain the template and corresponding constraint on it at MicroBooNE was the following~\cite{MicroBooNE:2018vnm}:
\begin{enumerate}
    \item The MiniBooNE Monte Carlo sample of true $\nu_e$ charged-current (CC) interactions is used to construct the response matrix $\mathcal{A}_{i\alpha} \equiv P\mathrm{(Reconstructed~in~}i\mathrm{|generated~in~}\alpha\mathrm{)}$, where $i$ refers to the reconstructed energy bin $(E_\nu^{\rm QE})_i$, and $\alpha$ refers to the true energy bin $(E_\nu^{\rm true})_\alpha$.
    \item The unfolded intrinsic $\nu_e$ CC interaction rate in true energy space, $u_\alpha$, is obtained via the D'Agostini iterative approach~\cite{DAgostini:1994fjx}.
    The regularization parameter was chosen to (1) minimize the variance of the unfolded spectra, (2) minimize the bias of the unfolded spectra, and (3) produce an expected event rate in reconstructed energy space, which is statistically consistent with the observed MiniBooNE data.
    \item The ratio of the unfolded event rate $u_\alpha$ and the central value MiniBooNE Monte Carlo prediction are taken as weights in true neutrino energy space, which are then applied to true $\nu_e$ CC interactions in the MicroBooNE simulation to produce the LEE model prediction.
\end{enumerate}

\begin{figure}[t]
    \centering
    \includegraphics[width=0.49\textwidth]{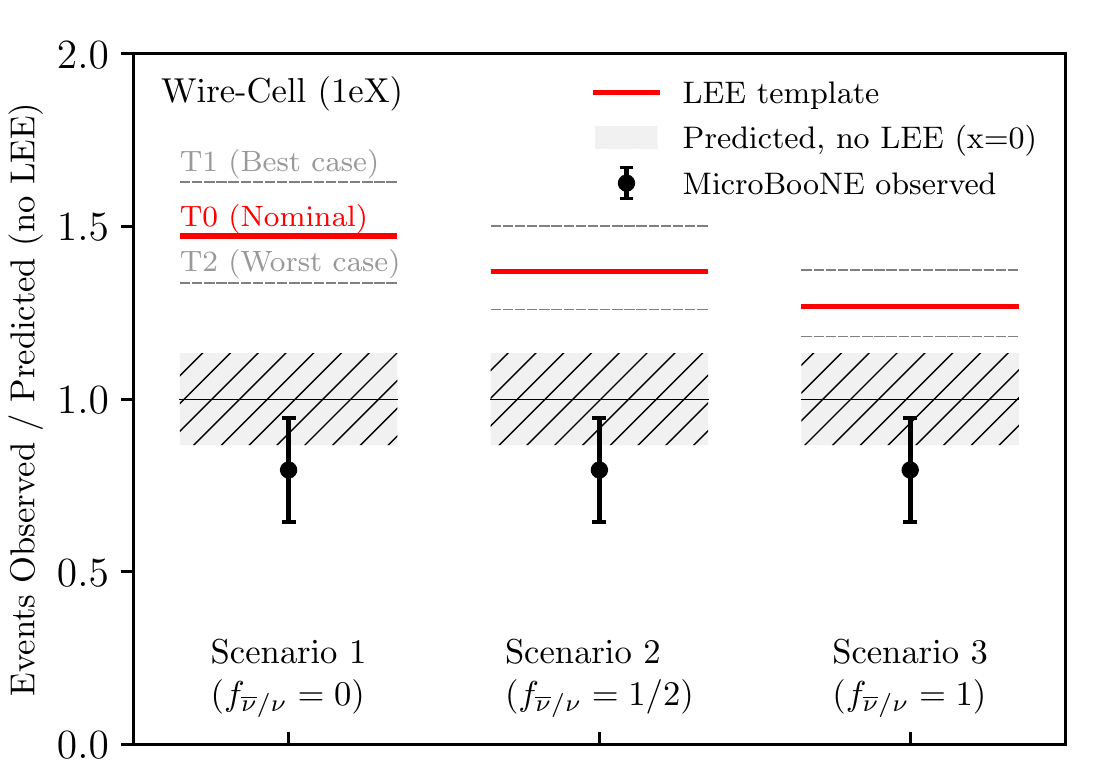}
    \caption{
    The ratio between observed and predicted LEE events at MicroBooNE in the three antineutrino-neutrino LEE scenarios.
    The grey bands show the uncertainty in the prediction in the absence of a LEE ($x = 0$), and in red, we show the prediction for the LEE with ($x=1$) in the three scenarios.
    From left to right, we show \emph{Scenario 1} ($f_{\overline{\nu}/\nu} = 0$), \emph{Scenario 2} ($f_{\overline{\nu}/\nu} = \frac{1}{2}$), and \emph{Scenario 3} ($f_{\overline{\nu}/\nu} = 1$).
    The fainter dashed grey lines represent the LEE prediction for the two alternative templates of Ref.~\cite{Arguelles:2021meu}.
    For the assumptions and methodology behind our analysis, see \cref{sec:analysis} and \cref{app:details}.
    \label{fig:scenarioresults}
    }
\end{figure}

The two MicroBooNE analyses with the highest sensitivity to the LEE were the Deep-Learning Based Analysis (DL)~\cite{MicroBooNE:2021bcu} and the Wire-Cell Analysis (WC)~\cite{MicroBooNE:2021nxr}.
In brief, the DL analysis looked for an exclusive sample of $\nu_e$ CC interactions with one electron and one proton in the final state ($1e1p$).
The reconstruction chain relied on two novel LArTPC-specific deep learning algorithms, \textsc{SparseSSNet}~\cite{MicroBooNE:2020yze} and \textsc{MPID}~\cite{MicroBooNE:2020hho}, to isolate these $1e1p$ events.
The kinematics of the electron and proton were required to be consistent with CC quasi-elastic scattering to reduce systematic uncertainties on the interaction cross-section.
This resulted in a signal sample with a large signal-to-background ratio but comparatively low statistics, with 25 events passing the full selection.
In the $E_{\nu/\overline{\nu}} < 400$~MeV range, because the DL analysis required a lepton-proton vertex, the selected events were almost entirely due to neutrino interactions rather than antineutrino interactions.
The energy distribution of DL-selected events was fit to the Standard Model prediction plus the MiniBooNE-based LEE model with floating normalization.
Based on this fit, the DL analysis limited the content of the MiniBooNE LEE to $<38\%$ $\nu_e$ interactions at $2\sigma$.
Since the DL analysis is not sensitive to antineutrinos, this result implies that more than $68\%$ of the excess can be due to $\overline{\nu}_e$ or other unrelated non-neutrino events.

\begin{figure}[t!]
    \centering
    \includegraphics[width=0.435\textwidth]{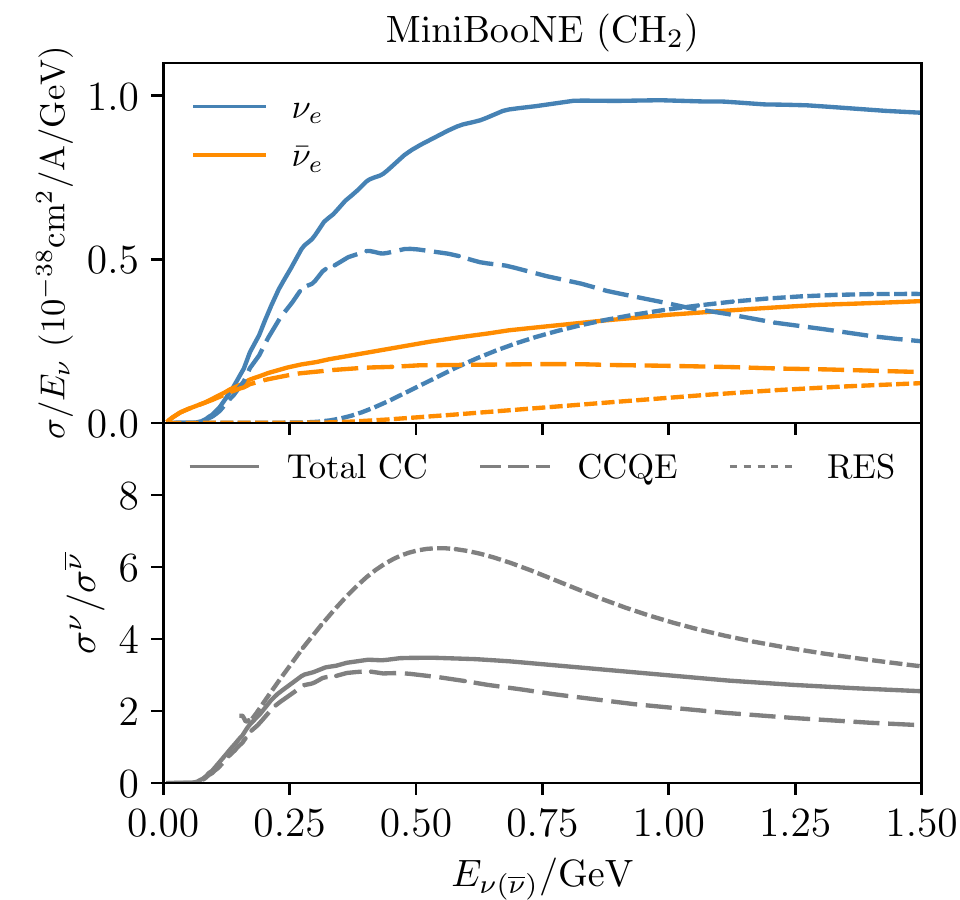}
    \includegraphics[width=0.435\textwidth]{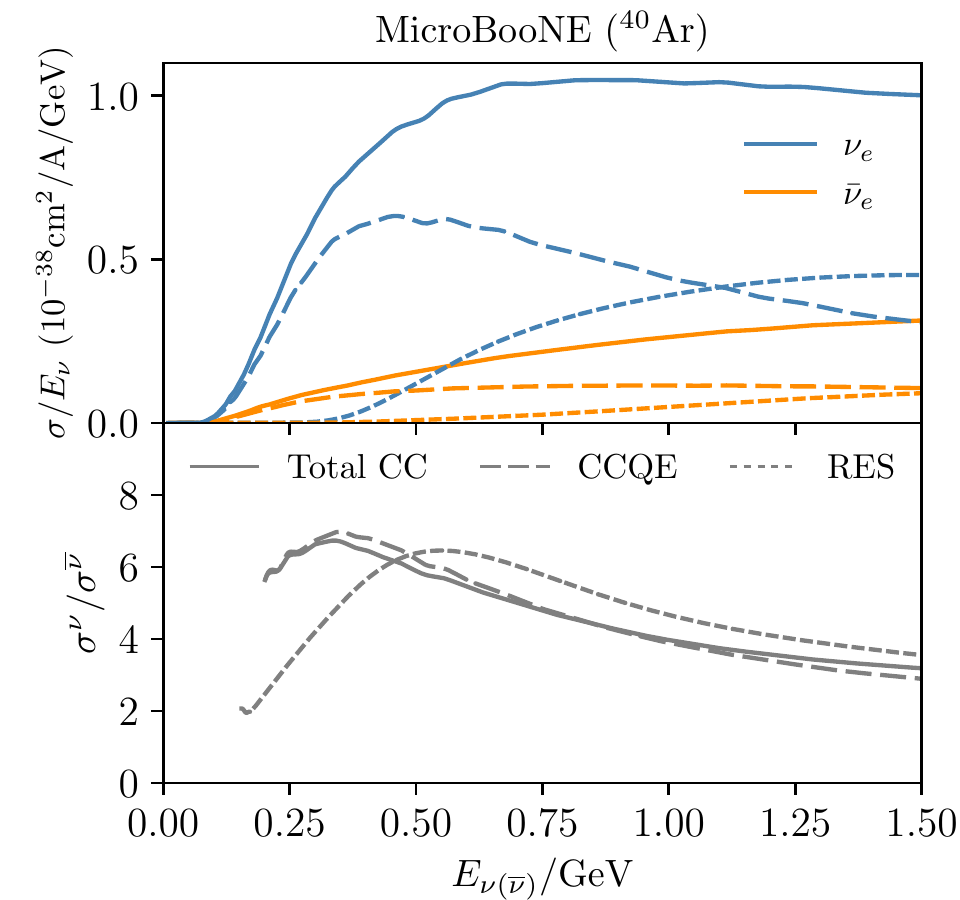}
    \includegraphics[width=0.435\textwidth]{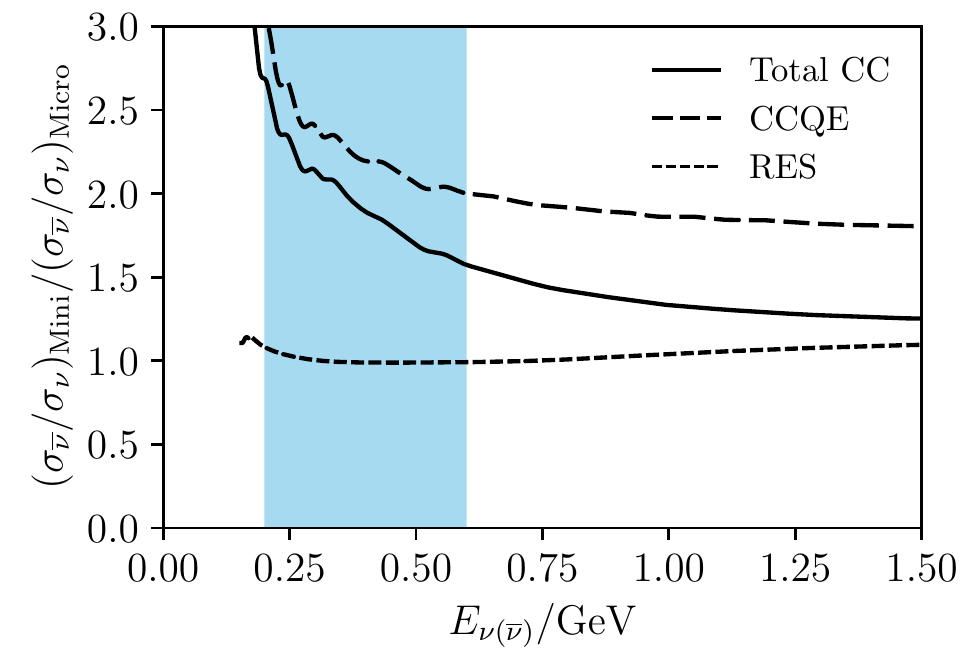}
    \caption{
    The top and middle panels show the total $\nu_e$ and $\overline{\nu}_e$ cross sections and their ratios for CH$_2$ and $^{40}$Ar, respectively.
    The CCQE and resonant cross sections are shown as dashed lines.
    In the bottom panel, we show the ratio between the MiniBooNE and MicroBooNE $\nu/\overline{\nu}$ ratios.
    The blue region indicates the region of the MiniBooNE LEE.
    \label{fig:cross_sections}}
\end{figure}

The WC analysis looked for an inclusive sample of $\nu_e$ CC interactions with one electron and anything else in the final state ($1eX$).
The namesake Wire-Cell algorithm~\cite{MicroBooNE:2021ojx} was used to identify three-dimensional space points of charge within each MicroBooNE image, which were then clustered and analyzed using a series of pattern recognition algorithms to isolate $1eX$ events.
The WC analysis allowed for single-lepton events as well as multi-prong vertices.
Thus, in the $E_{\nu/\overline{\nu}} < 400$~MeV range, soft as well as hard scattering processes dominate.
For WC, antineutrino events could contribute, although the missing energy due to the neutron would lead to an underestimate of the antineutrino energy.
Fitting the Standard Model prediction plus the MiniBooNE-based LEE model with floating normalization to the WC selected events, MicroBooNE set a $2\sigma$ upper limit of 50.2\% electron-flavor content in the LEE~\cite{MicroBooNE:2021nxr}.
In quoting this result, the collaboration assumed that the Wire-Cell analysis is not sensitive to the portion of the excess that is caused by sources other than an excess of electron neutrinos.

In the remainder of this article, we explore how the constraints above change when the assumption on the neutrino-antineutrino composition of the LEE is allowed to vary.
While the design of the DL analysis makes it insensitive to the content of antineutrinos in the LEE, it can still, in principle, be observed by the WC analysis.
As we will show, however, due to the fact that antineutrino-argon cross sections are much smaller than those of antineutrino-CH$_2$ for the LEE energies, the entire suite of MicroBooNE analyses turns out to be significantly less effective in constraining an excess of antineutrinos.
This is shown in \cref{fig:scenarioresults}, which indicates that the number of excess events predicted in the lowest energy region of the WC analysis decrease as the $\overline{\nu}_e$ content of the LEE increases from \emph{Scenario 1} to \emph{Scenario 3}.

The rest of this article proceeds as follows.
First, in \cref{sec:xsecs}, we compare the neutrino versus antineutrino interaction rates in the MiniBooNE and MicroBooNE targets, CH$_2$ and Ar, respectively.
We then generalize the unfolding of the MiniBooNE LEE to scenarios with a mixture of neutrino and antineutrino fluxes in \cref{sec:analysis}.
We apply our procedure to three assumptions on the neutrino-antineutrino composition of the LEE event rate:
100\% $\nu_e$, a 50\%-50\% mix, and 100\% $\overline \nu_e$.
Finally, we discuss the implications of these scenarios for the MicroBooNE analyses in \cref{sec:impact} and discuss the next steps in identifying sources of $\overline \nu_e$ in the MiniBooNE LEE in \cref{sec:discussion}.

\section{Neutrino and Antineutrino interactions in the LEE}\label{sec:xsecs}

In this section, we discuss the interaction cross sections for electron-neutrinos and electron-antineutrinos in the target material of the MiniBooNE and MicroBooNE detectors.
Their impact in the MicroBooNE analyses is discussed in \cref{sec:impact}.

At the fundamental level, the interaction cross sections of neutrinos and antineutrinos are different due to the sign of the axial-vector component.
For CCQE scattering on free nucleons, the difference can be expressed quite simply as
\begin{equation}\label{eq:xsec_diff}
    \frac{1}{\sigma_0}\left(\frac{\dd \sigma^\nu}{\dd y} - \frac{\dd \sigma^{\overline \nu}}{\dd y}\right) = y \left(1 - \frac{y}{2}\right) (F_1 + F_2)F_A,
\end{equation}
where $\sigma_0 \equiv \frac{G_F^2 |V_{ud}|^2 M E_\nu}{\pi}$ and $y = 1 - E_\ell/E_\nu$ is the inelasticity parameter.
In addition to the dependence on $y$, the above expression has an implicit dependence on the kinematics through the nucleon form factors $F_i \equiv F_i(Q^2)$, where $Q^2 = 2 E_\nu M y$ is the momentum exchange with the nucleon.
\Cref{eq:xsec_diff} constitutes the interference between axial-vector, $F_A \propto g_A$, and vector pieces of the amplitude.
This interference is destructive for antineutrinos but constructive for neutrinos.
Interestingly, \cref{eq:xsec_diff} leads to a preference for lower momentum transfer and thus more forward scattering angles of the final state lepton in the antineutrino case compared to the neutrino case, a behavior that is in better agreement with the forward-peaked nature of the MiniBooNE excess~\cite{MiniBooNE:2020pnu}.

For isoscalar targets, and in the absence of thresholds, $\sigma_{\nu} > \sigma_{\overline\nu}$.
At high energies, the ratio asymptotes to a factor of $\sim 1/2$.
At low energies, however, it can vary significantly due to threshold, nuclear, and binding energy effects.

\subsection{Cross sections at MiniBooNE and MicroBooNE}

The composition of the mineral oil in MiniBooNE is CH$_2$, providing six bound neutrons for neutrino CC interactions, and six bound and two free protons for antineutrino CC interactions.
In MicroBooNE, the argon nuclear targets provide 22 bound neutrons for neutrino CC interactions and 18 bound protons for antineutrinos.
These non-isoscalar materials enhance the number of nucleon targets for antineutrino CCQE at MiniBooNE by 33\% and suppress them at MicroBooNE by 10\%.

However, an even stronger effect is at play in the energy region of the LEE: the separation energy of protons and neutrons inside the argon nucleus.
As opposed to the isoscalar $^{12}$C nucleus, the non-isoscalar $^{40}$Ar nucleus contains protons that are more strongly bound than neutrons, and, therefore, require more energy to be knocked out by the CC interactions of antineutrinos.
This effect is important at small neutrino energies, where the center of mass energy is comparable with the nuclear binding energies.
In addition, the Pauli blocking of neutrons in argon is more significant than in carbon due to its size.

We show a comparison of the total and exclusive cross sections for neutrino and antineutrino cross sections on carbon and argon in \Cref{fig:cross_sections}, obtained from \textsc{GENIE\,v3.02.00}~\cite{Andreopoulos:2015wxa,GENIE:2021npt}.
While this ratio is similar for argon and CH$_2$ at high energies, it is significantly different at lower energies, varying by factors larger than two in the energy region of the LEE.
For the same event rate at MiniBooNE, this implies that MicroBooNE would see fewer events if antineutrinos induced those events rather than neutrinos.

We have elected to use \textsc{GENIE\,v3.02.00}, as it is the most up-to-date GENIE public release at the time of this study.
Other versions do exist--for example, the $CC0\pi$ MicroBooNE tune of \textsc{GENIE\,v3.00.06} presented in Ref.~\cite{MicroBooNE:2021ccs}.
While it would certainly be interesting to investigate the antineutrino hypothesis within the context of the MicroBooNE GENIE tune, it is not publicly available to our knowledge.
Therefore, we rely on \textsc{GENIE\,v3.02.00} for this study and leave the consideration of alternative neutrino event generators to future work.

To quantify the effect of an antineutrino component in the LEE, we should also consider the kinematics of the neutrino- and antineutrino-induced leptons.
Since the interference term between vector and axial components, shown in \Cref{eq:xsec_diff} for CCQE, is proportional to the inelasticity parameter $y = 1 - E_e/E_\nu$, when it contributes constructively, it leads to a preference for larger $y$, and, therefore, lower-energy leptons.
This is the case for neutrino-induced reactions.
In the case of antineutrinos, the interference is destructive, leading to a preference for smaller $y$, and, therefore, higher-energy leptons.
As we show below, in the context of the LEE, this implies that to reproduce the observed excess of events in \emph{Scenarios 2} and \emph{3}, the flux excess of antineutrinos would require a lower mean energy than the corresponding flux excess of neutrinos in \emph{Scenario 1}.

Note that this also implies that the impact of nuclear physics on the total cross sections, especially the suppression of low-$Q^2$ configurations, is different between neutrinos and antineutrinos.
This dependence on the lepton kinematics and hadronic energy means that the unfolding procedure adopted by MicroBooNE ought to be modified before applying it to the antineutrino hypothesis.

\section{The MicroBooNE unfolding-based template analysis} \label{sec:analysis}

To unfold the MiniBooNE excess under antineutrino-based explanations of the LEE, we follow the procedure outlined by the MicroBooNE collaboration in Ref.~\cite{MicroBooNE:2018vnm}.
Specifically, we use the D'Agostini iterative approach~\cite{DAgostini:1994fjx} to unfold the observed MiniBooNE data, where the prediction $u^k_\alpha$ in true energy bin $\alpha$ at iteration $k$ is given by
\begin{equation}
    u_\alpha^k = \sum_{i=1}^{n_r} M^{k-1}_{i \alpha} d_i,
\end{equation}
where the sum goes over each of the $n_r$ reconstructed energy bins, and $d_i$ denotes the observed data in reconstructed bin $i$.
The matrix $M^k_{i \alpha }$ is defined by
\begin{equation}{\label{eq:unfolding_matrix}}
    M^k_{i \alpha } =
    \frac{
    A_{i\alpha} u_\alpha^k
    }
    {
    \epsilon_\alpha \sum_{\beta=1}^{n_t} A_{i\beta} u^k_\beta
    },
\end{equation}
where the response matrix $A_{i\alpha}$ is given by
\begin{equation}
    A_{i\alpha} = {\rm P}({\rm reconstructed~in~}i|{\rm generated~in~}\alpha),
\end{equation}
and $\epsilon_\alpha \equiv \sum_i A_{i \alpha}$ is the reconstruction efficiency in true energy bin $\alpha$.

\subsection{Introducing an antineutrino component}

\begin{figure}[th!]
    \centering
    \includegraphics[width=0.49\textwidth]{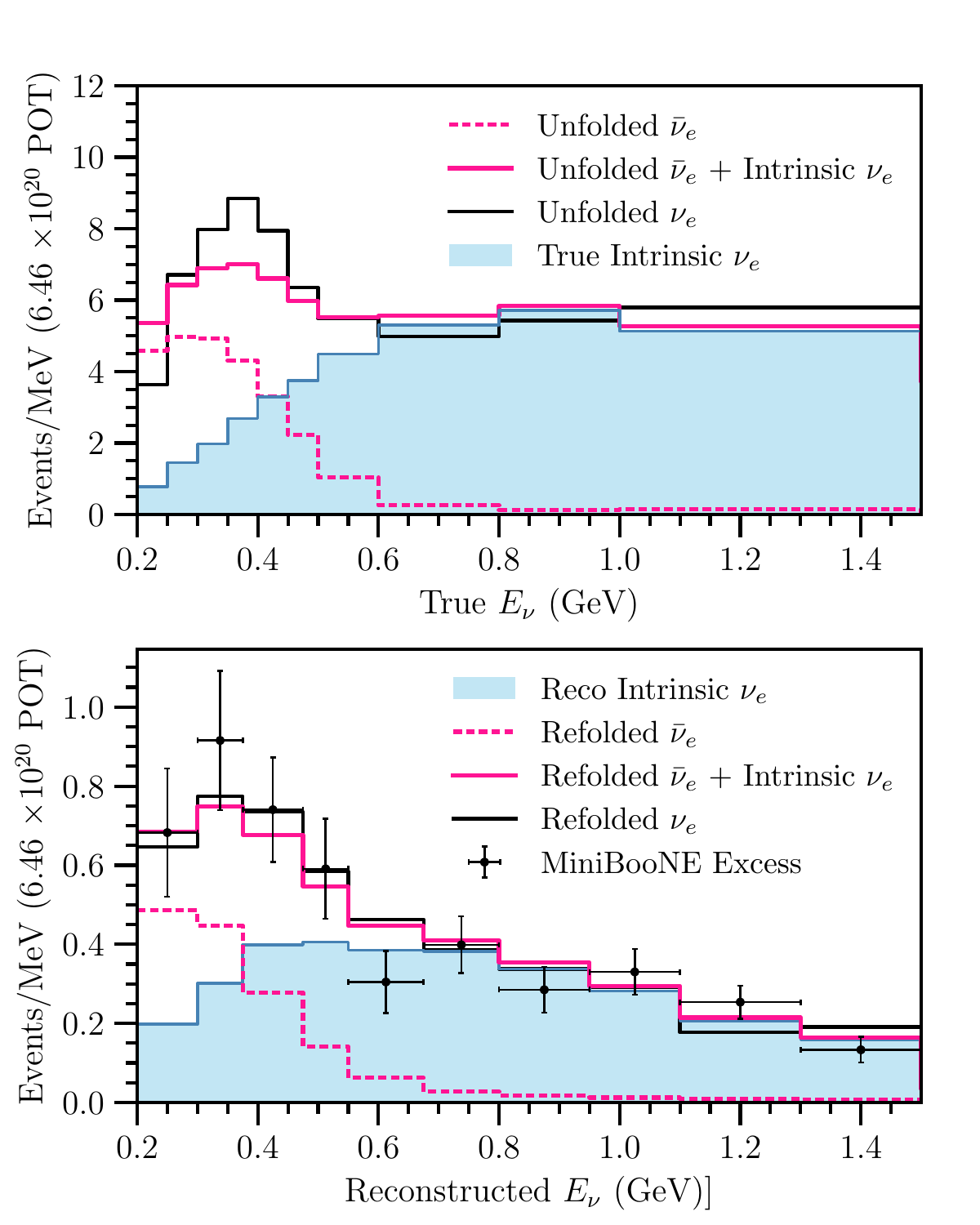}
     \caption{
     In the top panel we show the intrinsic $\nu_e$ background and the unfolded $\nu_e$ and $\overline{\nu}_e$ templates obtained via D'Agostini's unfolding method.
     The $\nu_e$ unfolding procedure begins with the intrinsic $\nu_e$ event rate as an initial guess, while the $\overline{\nu}_e$ unfolding procedure begins with a flat distribution in the true neutrino energy.
     In the bottom panel, we show the re-folded prediction in MiniBooNE for both the unfolded $\nu_e$ hypothesis and unfolded $\overline{\nu}_e$ plus intrinsic $\nu_e$ hypothesis, compared with the excess data for $6.46 \times 10^{20}$ POT~\cite{MicroBooNE:2018vnm}.
     \label{fig:unfolded_templates}
     }
\end{figure}

\begin{figure*}[th!]
\centering
\includegraphics[width=0.49\textwidth]{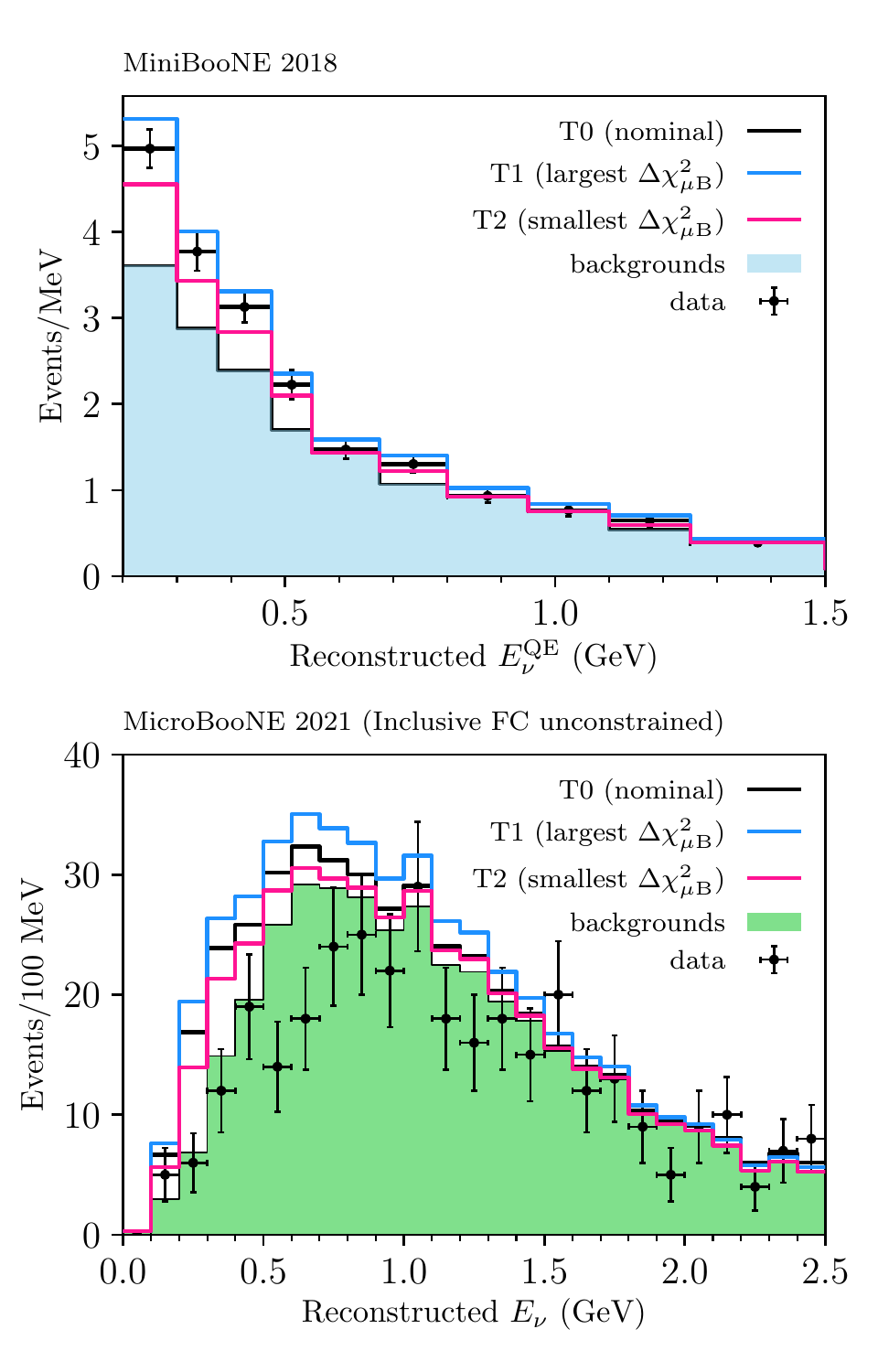}
\includegraphics[width=0.49\textwidth]{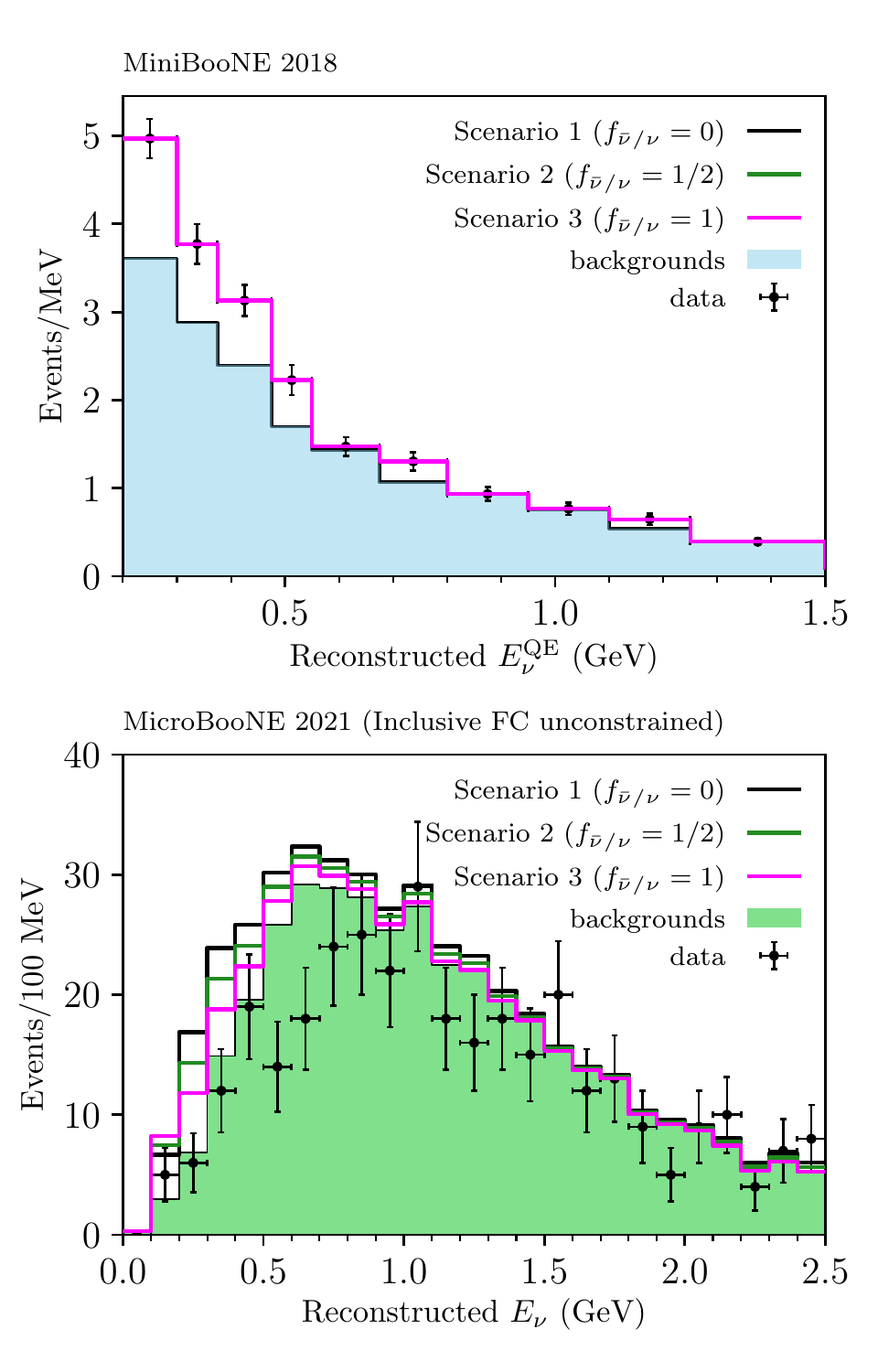}
\caption{The MiniBooNE and MicroBooNE spectra in reconstructed neutrino energy.
On the left panels, we show three different template choices from Ref.~\cite{Arguelles:2021meu} and their corresponding prediction at MicroBooNE, assuming no antineutrinos, $\f = 0$.
On the right panels, we show how the three different assumptions for the antineutrino composition of the excess impact the nominal template (T0) prediction in MicroBooNE.
In MiniBooNE, the three different scenarios of antineutrino compositions look exactly the same, by definition.
\label{fig:templates}}
\end{figure*}

The neutrino and antineutrino response matrices will differ in MiniBooNE, as the $\dd\sigma/\dd y$ distribution prefers smaller (larger) values of $y$ for neutrino (antineutrino) CCQE scattering.
We calculate the antineutrino response matrix in MiniBooNE by simulating $\overline{\nu}_e$ charged-current interactions in CH$_2$ using \textsc{GENIE\,v3.02.00}.
This formalism is unable to account for the MiniBooNE reconstruction efficiency; thus, we instead estimate $R_{i\alpha} \equiv A_{i\alpha}/\epsilon_\alpha$.

As Cherenkov detectors are only sensitive to the final state lepton, MiniBooNE uses $E_\nu^{\rm QE}$ ($E_{\overline{\nu}}^{\rm QE}$) to reconstruct $\nu_e$ ($\overline{\nu}_e$) energies, given by the expressions
\begin{equation} \label{eq:enuqe}
    E_\nu^{\rm QE} = \frac{
    2(M'_n)E_\ell - ((M'_n)^2 + m_\ell^2 - M_p^2)
    }
    {
    2[(M'_n) - E_\ell + \sqrt{E_\ell^2 - m_\ell^2}\cos\theta_\ell]
    },
\end{equation}
\begin{equation}
    E_{\overline{\nu}}^{\rm QE} = \frac{
    2(M'_p)E_\ell - ((M'_p)^2 + m_\ell^2 - M_n^2)
    }
    {
    2[(M'_p) - E_\ell + \sqrt{E_\ell^2 - m_\ell^2}\cos\theta_\ell]
    }.
\end{equation}
Here, $M_n$ and $M_p$ are the neutron and proton mass, $E_\ell$, $m_\ell$, and $\theta_\ell$ are the lepton energy, mass, and scattering angle, and $M'_{(n/p)} \equiv M_{(n/p)} - E_B$, where the nucleon binding energy $E_B$ is fixed to 34 (30) MeV for neutrinos (antineutrinos).
It is important to emphasize that no matter the origin of the underlying event ($\nu_e$ or $\overline{\nu}_e$), interactions are always reconstructed using $E_\nu^{\rm QE}$ ($E_{\overline{\nu}}^{\rm QE}$) for data taken in neutrino (antineutrino) mode.

We consider only neutrino mode data in this study, as this is directly comparable to the MicroBooNE neutrino mode data.
We approximate $R_{i \alpha}$ by marginalizing over $E_\nu^{\rm QE,true}$, using \textsc{GENIE\,v3.02.00} to generate the truth-level final state kinematic distributions of the $e^+$ which appear in \cref{eq:enuqe}.
The details of this calculation are given in \cref{app:response_matrix}.
We separately approximate the $\overline{\nu}_e$ detection efficiency in MiniBooNE by using the provided detection efficiency as a function of electron energy~\cite{MiniBooNE_eff}.
The details of the efficiency calculation are given in \cref{app:approximations}.

Armed with our calculation for $R_{i \alpha} = A_{i\alpha}/\epsilon_\alpha$ and $\epsilon_\alpha$, we can perform the unfolding procedure using \cref{eq:unfolding_matrix}.
This produces a prediction for the antineutrino interaction rate in MiniBooNE as a function of true antineutrino energy, hereafter denoted $u^{\rm MB}_\alpha$, which we will use to predict a signal in MicroBooNE.
In the top panel of \cref{fig:unfolded_templates}, we show our unfolded $\overline{\nu}_e$ prediction considering the first $6.46 \times 10^{20}$ POT of MiniBooNE data--the same dataset used in Ref.~\cite{MicroBooNE:2018vnm}.
One can see that the unfolded $\overline{\nu}_e$ template peaks at lower (anti)neutrino energy compared with the unfolded $\nu_e$ template.
In the bottom panel of \cref{fig:unfolded_templates}, we re-fold the unfolded excess templates back through the MiniBooNE reconstruction.
Both the neutrino and antineutrino re-folded predictions are in agreement with the MiniBooNE data.

Once we have an unfolded $\overline{\nu}_e$ MiniBooNE prediction $u^{\rm MB}_\alpha$, we can fold that prediction into MicroBooNE.
In contrast with~\cref{fig:unfolded_templates}, here we use the $12.84 \times 10^{20}$ POT neutrino mode dataset presented in MiniBooNE's 2018 result~\cite{MiniBooNE:2018esg} for the unfolding process.
This is the same dataset used by the MicroBooNE collaboration to calculate their LEE template~\cite{MicroBooNE:2021nxr}.
Note that the unfolded spectrum represents the \textit{interaction rate} inside MiniBooNE -- thus, we must scale by the ratio of cross sections in Ar and CH$_2$ when going to MicroBooNE.
The predicted MicroBooNE event rate $\mu^{\mu \rm{B}}_i$ in $E_{\overline{\nu}}^{\rm reco}$ bin $i$ is given by
\begin{equation}
\mu^{\mu \rm{B}}_i = \sum_\alpha \epsilon_\alpha R_{i\alpha}  u^{\rm MB}_\alpha \frac{\sigma_{{\rm Ar}}((E_{\overline{\nu}}^{\rm true})_\alpha)}{\sigma_{{\rm CH}_2}((E_{\overline{\nu}}^{\rm true})_\alpha)},
\end{equation}
where $\epsilon_\alpha$ and $R_{i\alpha}$ now denote the MicroBooNE $\overline{\nu}_e$ detection efficiency and MicroBooNE response matrix, respectively.

The $\overline{\nu}_e$ response matrix calculation in MicroBooNE is similar to the MiniBooNE calculation.
The main difference is that MicroBooNE, being a LArTPC, is able to perform a calorimetric energy reconstruction.
The reconstructed energy in MicroBooNE is given by $E_\nu^{\rm Cal} = \sum_j (T_j^{\rm reco} + m_j + B_j),$ where $T_j^{\rm reco}$, $m_j$, and $B_j$ denote the observed kinetic energy, rest mass, and binding energy associated with the $j$'th reconstructed final state particle.
The binding energy $B_i$ is taken to be 8.6 MeV for protons and zero for everything else.
In the case of $\overline{\nu}_e$ CCQE scattering, MicroBooNE will only reconstruct the final state $e^+$.
The reconstructed energy $E_{\overline{\nu}}^{\rm reco}$ is then simply $E_{\overline{\nu}}^{\rm Cal,reco} = T_{e^+}^{\rm reco} + m_{e^+}$.
Note that this will lead to an under-estimation bias in the reconstructed $\overline{\nu}_e$ energy due to the invisible neutron.
In order to calculate $R_{i \alpha}$ in MicroBooNE, we marginalize over the $T_{e^+}^{\rm true}$ distribution generated using \textsc{GENIE\,v3.02.00}.
The details of this calculation are given in \cref{app:response_matrix}.

Evaluating the detection efficiency of $\overline{\nu}_e$ interactions in MicroBooNE as a function of $E_{\overline{\nu}_e}$ is more complicated than the MiniBooNE case.
MicroBooNE is not a spherically symmetric detector, thus both the electron direction and energy will impact the detection efficiency.
The MicroBooNE reconstruction also relies non-trivially on hadronic information in addition to leptonic information for its nominal $\nu_e$ analyses, including the inclusive analysis studied here.
Ignoring final state interactions, there will be no hadronic information in $\overline{\nu}_e$ CCQE interactions.
It is also possible for $\nu_e$ CCQE interactions to occur without hadronic activity if the energy of the final state proton is below the reconstruction threshold; however, this is considerably different than the $\overline{\nu}_e$ case, as the final state neutron can carry away an arbitrary amount of energy without being reconstructed.
Given these complications, a detailed estimation of the MicroBooNE $\overline{\nu}_e$ efficiency is out of the scope of this paper.
We instead conservatively assume the $\overline{\nu}_e$ efficiency in MicroBooNE to be the same as the reported
$\nu_e$ efficiency for a given true (anti)neutrino energy.
In a realistic scenario, the $\overline{\nu}_e$ efficiency is likely smaller due to the lack of hadronic information; thus, one can interpret the MicroBooNE $\overline{\nu}_e$ prediction derived here as an upper bound.

\section{The impact on the MicroBooNE template analysis}\label{sec:impact}

\begin{figure}[t]
\centering
\includegraphics[width=0.49\textwidth]{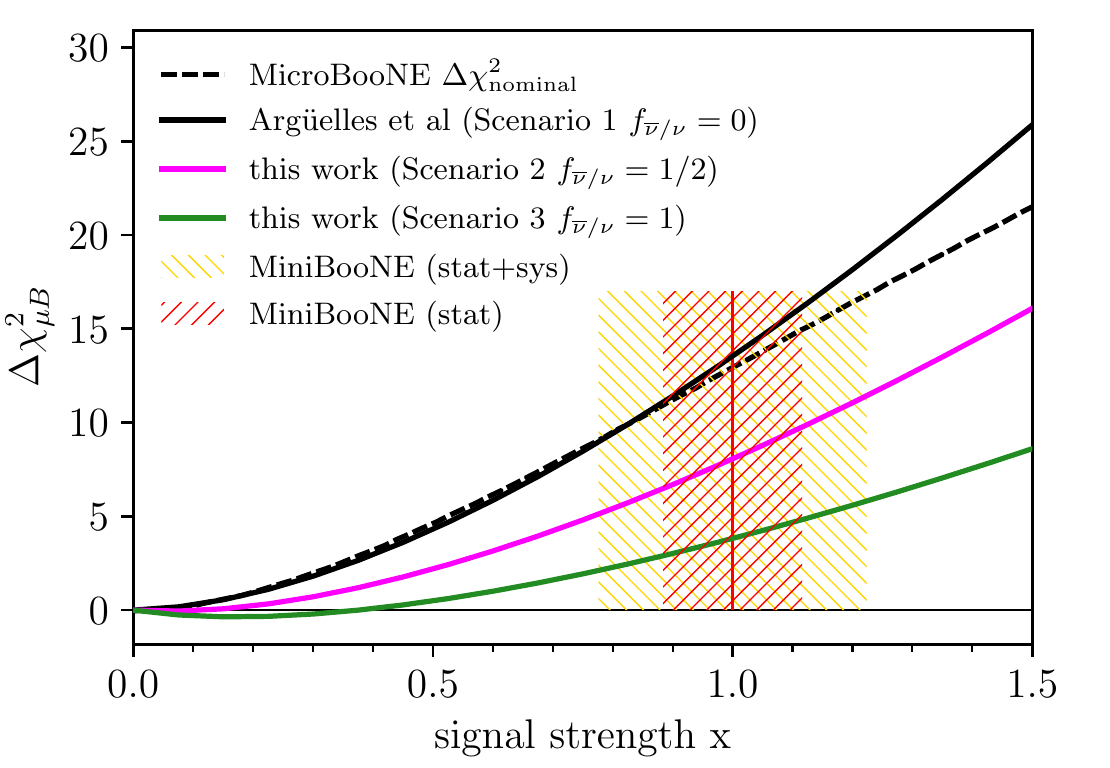}
\caption{The MicroBooNE $\Delta \chi^2$ for the inclusive $1eX$ analysis as a function of the signal strength $x$.
The black solid line shows the $\Delta \chi^2$ for the nominal template, following the calculation of Ref.~\cite{Arguelles:2021meu}.
There is good agreement with the official MicroBooNE curve, shown in dashed black.
In solid green and violet lines, we show the corresponding curves for \emph{Scenarios 2} and \emph{3} for the neutrino-antineutrino fractions of the LEE, respectively.
\label{fig:chi2_sigstrength}}
\end{figure}

The right panels of \cref{fig:templates} show the prediction in MicroBooNE under the three different hypotheses for the antineutrino content in the LEE outlined in \cref{sec:intro}: \emph{Scenario 1} ($\f = 0$), \emph{Scenario 2} ($\f = 1/2$), and \emph{Scenario 3} ($\f = 1$).
As can be seen, the LEE template prediction in MicroBooNE from the unfolded MiniBooNE excess decreases in general with $\f$, the antineutrino fractional contribution to the excess.
Thus, as expected, MicroBooNE is less sensitive to antineutrino-based explanations of the MiniBooNE excess.
This is quantified in \cref{fig:chi2_sigstrength}, which shows the test statistic of the Wire-Cell analysis, $\Delta \chi^2_{\mu B}(x) \equiv \chi^2_{\mu B}(x) - \chi^2_{\mu B}(x=0)$, as a function of the signal strength scaling parameter $x$ introduced in Ref.~\cite{MicroBooNE:2021rmx}.
As shown in \cref{tab:numerical_results_sig_strength}, the exclusion power drops significantly as the predicted antineutrino content becomes larger.
Specifically, as $\f$ increases from 0 to 0.5, the test statistic $\Delta \chi^2_{\mu B}(x=1.0)$ falls from \dchisqnomNone~to \dchisqnomHalf.
For $\f = 1.0$, $\Delta \chi^2_{\mu B}(x=1.0) = \dchisqnomFull$.
At around $x=0.2$, the $\f = 1.0$ case predicts a slightly negative $\Delta \chi^2_{\mu B}$, implying a minor improvement with respect to the nominal BNB prediction.
This is most likely caused by the small excess observed in the lowest energy bin of the MicroBooNE analysis, as shown in the lower-right panel of \cref{fig:templates}.

Assuming Wilks' theorem~\cite{Wilks:1938dza} with one degree-of-freedom, the critical $\Delta \chi^2_{\mu B}$ value at the 95.45\% ($2\sigma$) confidence level is $\Delta \chi^2_{\mu B} = 4$.
We use this to calculate $2\sigma$ upper limits on the signal scaling parameter $x$ in \emph{Scenarios} 1, 2, and 3, considering the test statistic $\Delta \chi^2_{\mu B}(x) = \chi^2(x) - \min_x \{ \chi^2(x)\}$.
These $2\sigma$ upper limits are shown in \cref{tab:numerical_results_sig_strength}.
From \cref{tab:numerical_results_sig_strength}, it is evident that MicroBooNE is much less sensitive to \emph{Scenario 2} than \emph{Scenario 1}, and is essentially insensitive to \emph{Scenario 3}.

\begin{figure}[t]
\centering
\includegraphics[width=0.49\textwidth]{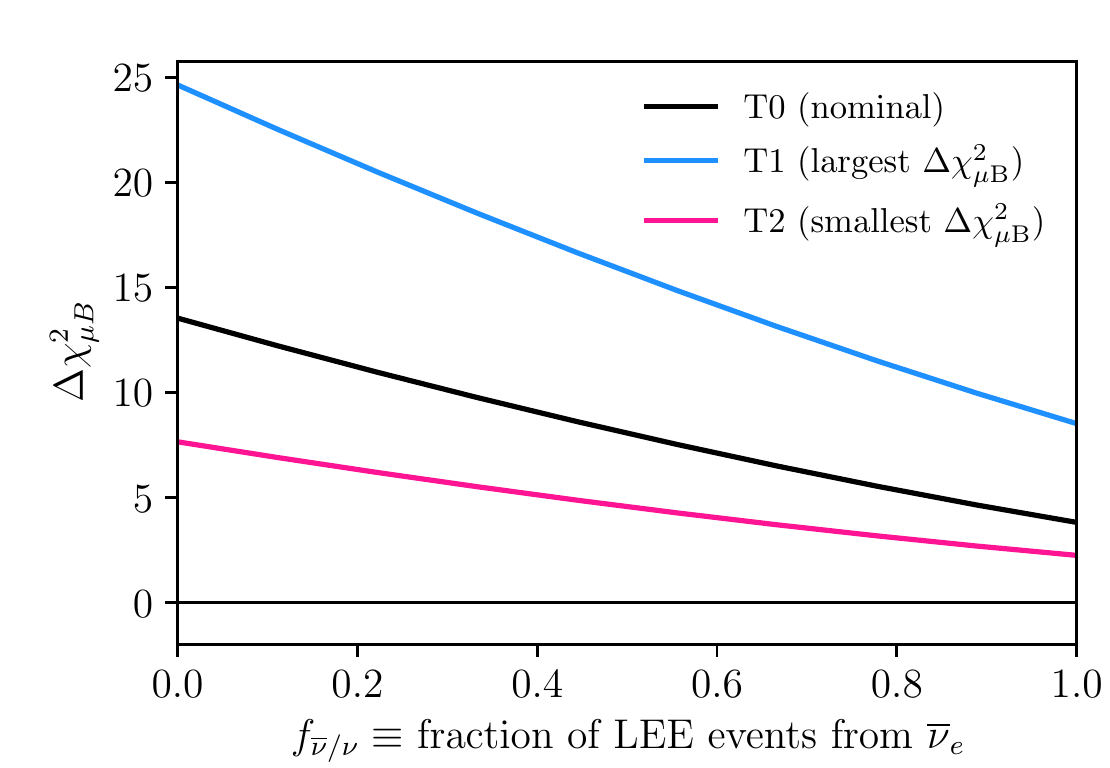}
\caption{The MicroBooNE $\Delta \chi^2$ for the inclusive $1eX$ analysis as a function of the antineutrino-neutrino fraction of the excess, $\f$.
In black, we show the variation for the nominal template, while in blue and pink, we repeat the same exercise for the two templates shown in the left panels of \Cref{fig:templates}.
\label{fig:chi2_sigstrength_vs_frac}}
\end{figure}

In the left panels of \Cref{fig:templates}, we show three different templates for the MiniBooNE LEE: T0, T1, and T2.
Template T0 is the nominal unfolded template.
Templates T1 and T2, defined in Fig.~2 of Ref.~\cite{Arguelles:2021meu}, correspond to the most and fewest number of excess events obtained in the unfolding procedure while remaining consistent with the MiniBooNE excess at $p>80\%$.
In \cref{fig:chi2_sigstrength_vs_frac}, we show the $\Delta \chi^2_{\mu B}$ exclusion power of the MicroBooNE data as a function of the antineutrino fraction of the MiniBooNE excess, considering each template separately.
Here, $\Delta \chi^2_{\mu B}$ is defined with respect to the nominal BNB prediction, and we assume the nominal signal strength scaling for each template, i.e., $x=1.0$.
From \cref{fig:chi2_sigstrength_vs_frac}, one can clearly see that $\Delta \chi^2_{\mu B}$ decreases rapidly as the $f$ increases.
This is quantified in \cref{tab:numerical_results_f}, which reports the value of $\Delta \chi^2_{\mu B}$ for each template under \emph{Scenario 1} ($\f = 0$), \emph{Scenario 2} ($\f = 0.5$), and \emph{Scenario 3} ($\f = 1.0$).
Note that the results for the nominal template (T0) are the same as those reported in \cref{tab:numerical_results_sig_strength}.
Even for the most optimistic case (T1), the exclusion power is significantly suppressed for $\f = 1.0$.

We can define the test statistic $\Delta \chi^2_{\mu B}(\f) \equiv \chi^2_{\mu B}(\f) - \min_{\f \in [0,1]} \{ \chi^2_{\mu B}(\f) \}$ to calculate the $2\sigma$ lower limit on $\f$ for each template.
These lower limits are also reported in \cref{tab:numerical_results_f}; depending on the template they hover around $\f \sim 0.5$.
Thus if we take the MiniBooNE excess at face-value (i.e., restrict to $x=1.0$) and consider the nominal template (T0), the MicroBooNE results require at least $\percentage{\fLowerNom}$ of the MiniBooNE excess come from $\overline{\nu}_e$ events at the $2\sigma$~C.L.
Note that this statement relies on restricting ourselves to $x=1.0$--as shown in \cref{tab:numerical_results_sig_strength}, $\chi^2_{\mu B} (x=1.0) - \chi^2_{\mu B} (x=0) = 8.06$ for $\f = 0.5$, meaning that there is just under $3\sigma$ tension with the no-excess hypothesis ($x=0$) when attributing half of the MiniBooNE excess to $\overline{\nu}_e$ events.

So far, we have focused on the total $1eX$ sample, using the 7-channel fit of Ref.~\cite{MicroBooNE:2021nxr}.
However, as a consistency check, the Wire-Cell analysis has also performed an 11-channel fit, separating $\nu_\mu$ and $\nu_e$ CC events into samples with and without final state protons, $0pX\pi$ and N$pX\pi$.
Antineutrinos will contribute almost exclusively to the $0pX\pi$ sample, making it a purer sample of $\overline{\nu}_e$ LEE events.
The total number of $\nu_e$CC events is approximately even between the two samples, namely $259$ $0pX\pi$ and $298$ N$pX\pi$ events.
Given that the statistical and systematic uncertainties are larger for the $0pX\pi$ sample, and that it does not observe a deficit of events like that of the N$pX\pi$ sample ($N_{\rm data}/N_{\rm pred} = (1.00 \pm 0.08\text{ stat} \pm 0.21\text{ sys})$ versus $N_{\rm data}/N_{\rm pred} = (0.86 \pm 0.06\text{ stat} \pm 0.17\text{ sys})$), we do not expect that an 11-channel fit would qualitatively change our conclusions.

\renewcommand{\arraystretch}{1.25}
\begin{table}[t]
    \centering
    \begin{tabular*}{0.49\textwidth}{@{\extracolsep{\fill}}|c|c|c|c|}
     \hline
     & $\f=0.0$ & $\f=0.5$ & $\f=1.0$   \\
     \hline
     $\Delta \chi^2_{\mu B} (x=1.0)$ & $\dchisqnomNone$ & $\dchisqnomHalf$ & $\dchisqnomFull$ \\
     $2\sigma$ upper bound on $x$ & $\xUpperNone$ & $\xUpperHalf$ & $\xUpperFull$ \\
     \hline
    \end{tabular*}
    \caption{Statistical results from the MicroBooNE Wire-Cell analysis on the signal strength scaling parameter $x$, considering the three different scenarios outlined in \cref{sec:intro}. $\Delta \chi^2_{\mu B} (x=1.0)$ is reported with respect to the nominal BNB prediction, i.e., without any additional MiniBooNE-like excess.
    \label{tab:numerical_results_sig_strength}}
\end{table}

\begin{table}[t]
    \centering
    \begin{tabular*}{0.49\textwidth}{@{\extracolsep{\fill}}|c|c|c|c|}
     \hline
     Template & T0 & T1 & T2   \\
     \hline
     \multicolumn{4}{|c|}{\makebox[0pt]{Using $\Delta \chi^2_{\mu B}(\f) = \chi^2_{\mu B}(\f,x=1.0) - \chi^2_{\mu B} (x=0)$}} \\
     \hline
     $\Delta \chi^2_{\mu B} (\f=1.0)$ & $\dchisqnomFull$ & $\dchisqbestFull$ & $\dchisqworstFull$ \\
     $\Delta \chi^2_{\mu B} (\f=0.5)$ & $\dchisqnomHalf$& $\dchisqbestHalf$ & $\dchisqworstHalf$  \\
     $\Delta \chi^2_{\mu B} (\f=0.0)$ & $\dchisqnomNone$ & $\dchisqbestNone$ & $\dchisqworstNone$ \\
     \hline
     \multicolumn{4}{|c|}{\makebox[0pt]{Using $\Delta \chi^2_{\mu B}(\f) = \chi^2_{\mu B}(\f) - \min_{\f \in [0,1]} \{ \chi^2_{\mu B}(\f) \}$}} \\
     \hline
     $2\sigma$ lower bound on $\f$ & $\fLowerNom$ & $\fLowerBest$ & $\fLowerWorst$ \\
     \hline
    \end{tabular*}
    \caption{Statistical results from the MicroBooNE Wire-Cell analysis on the antineutrino fractional contribution to the MiniBooNE excess $\f$, considering the three different templates shown in the left panels of \cref{fig:templates}.
    For the first three rows, $\Delta \chi^2_{\mu B} (\f)$ is calculated as in \cref{fig:chi2_sigstrength_vs_frac}, while the last row considers the $\Delta \chi^2_{\mu B} (\f)$ definition given in the text to calculate $2\sigma$ lower limits on $\f$, assuming Wilks' theorem for 1 d.o.f.
    \label{tab:numerical_results_f}}
\end{table}

\section{Discussion}\label{sec:discussion}

Our main finding is that explanations of the MiniBooNE LEE involving a large contribution of wrong-sign electron-antineutrinos, from new physics or mis-modeling in the experimental simulation, remain viable.
The relative suppression of antineutrino cross sections in argon, the target material used by MicroBooNE, with respect to CH$_2$, the mineral oil target material used by MiniBooNE, means that MicroBooNE is much less sensitive to a low-energy excess of antineutrinos compared to neutrinos.
This motivates new strategies to measure the electron-antineutrino component of the BNB.

\begin{figure}[t]
    \centering
    \includegraphics[width=0.49\textwidth]{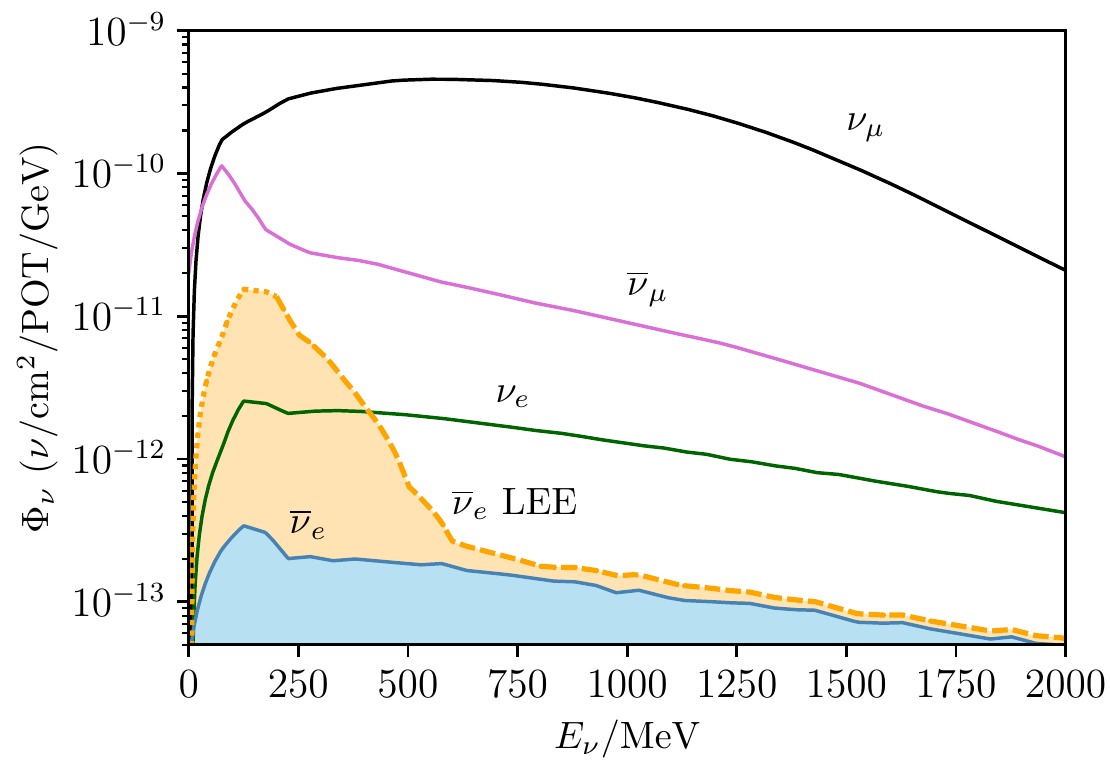}
    \caption{A smooth interpolation of the BNB neutrino fluxes in FHC mode as a function of energy.
    In light blue, we highlight the nominal prediction for the intrinsic $\overline\nu_e$~\cite{MiniBooNE:2008hfu}.
    In orange, we show the required excess $\overline{\nu}_e$ flux to explain the MiniBooNE low-energy excess in \emph{Scenario 3}, using the unfolded template of \cref{fig:unfolded_templates} (top).
    The dashed line is the result using the analysis range, $E_\nu > 200$~MeV.
    The dotted line in the $E_\nu < 200$~MeV region shows an extrapolation of the unfolded template, assuming it follows the same shape as the intrinsic $\overline{\nu}_e$ component.
    \label{fig:flux_excess}}
\end{figure}

The templates in the top panel of \cref{fig:unfolded_templates} unfolded under the hypothesis of an antineutrino-induced LEE indicate a flux excess that is even lower in energy than its neutrino-induced LEE counterpart.
As shown in \cref{fig:flux_excess}, such an excess would represent a significant deviation from the BNB model prediction for the flux of intrinsic $\overline{\nu}_e$ in neutrino mode~\cite{MiniBooNE:2008hfu}.
While no source for such hypothetical enhancement has been identified, this study provides a first glimpse into its energy dependence and relative rate.
In what follows, we discuss the implication of these findings for a few different antineutrino hypotheses.

\subsection{The BNB model}

The BNB flux model~\cite{MiniBooNE:2008hfu} predicts that the wrong-sign electron-neutrinos constitute a total of $0.05\%$ ($0.2\%$) of the total flux in neutrino (antineutrino) mode, arising primarily from charged and neutral kaons as well as secondary muons.~\footnote{The chirally-suppressed $\pi^- \to e^- \overline\nu_e$ ($\pi^+ \to e^+ \nu_e$) decays contribute at a much smaller fraction, namely $6.3 \times 10^{-6}$ ($1.6 \times 10^{-5}$) of the total neutrino (antineutrino) mode flux.}
An antineutrino explanation to the LEE requires a $10$ times larger flux of wrong-sign neutrinos than predicted in the BNB model.
Considering only the LEE region, $E_\nu < 600$~MeV, it requires a $25$ times larger flux.
Below we address the implications of such an enhancement under different hypotheses.

\paragraph{Wrong-sign pions} The early decays of forward-going wrong-sign pions in the BNB, $\pi^- \to \mu^- \overline\nu_\mu$, also appears as a peak at $E_\nu \lesssim 400$~MeV.
This process is associated with large uncertainties due to the lack of $\pi^-$ production data in the forward direction~\cite{MiniBooNE:2008hfu}.
The question then arises: could a large excess of wrong-sign pions in the low-energy region explain the MiniBooNE LEE?
The $\pi^- \to \overline\nu_e$ decays cannot be the only source of the excess, as it would be accompanied by an enormous (lower-energy) flux of $\pi^- \to \overline\nu_\mu$, exceeding the total neutrino flux below $600$~MeV by more than two orders of magnitude.
Another new source of $\overline\nu_e$ from wrong-sign pions are secondary muon decays, namely $\pi^- \to \mu^- \to \overline{\nu}_e$.
The BNB model predicts that neutrinos from secondary muons correspond to a $\sim 0.2\%$ fraction of the neutrinos from the parent pion.
To explain the LEE, this would require a $\sim500$ times larger $\pi^- \to \overline\nu_\mu$ flux, which is, again, not realistic.

\paragraph{Secondary muons}
Another logical possibility is that the primary neutrinos from wrong-sign pions are correctly modeled, but the subsequent decays of secondary muons are not.
This effect would have to account for a sizeable increase of the average energy of secondary-muon neutrinos, and, more importantly, would require a fifty-fold enhancement of the fraction between $\mu^-$ and $\pi^-$ neutrinos, bringing it to $10\%$.
Because of the short decay pipeline, it is difficult to conceive of a scenario where so many forward-going muons could contribute to the neutrino flux.
A forward-going muon produced at the target has a small probability $P \sim  50 \text{ m}/(\gamma_\mu c\tau^{0}_\mu)\sim 7.5\%/\gamma_\mu$ of decaying before hitting the beam absorber $\sim 50$~m downstream.
Here, $\gamma_\mu$ is the Lorentz boost of the muon, which can range from $\gamma_\mu = 2 - 10$ in the energy region of interest.
While the muons can penetrate the absorber and subsequent dirt, they will not all decay to produce a forward-going $\overline\nu_e$, so this possibility is also unrealistic.

\paragraph{Associated muon-neutrinos}
It is also reasonable to assume that a $\overline\nu_e$ excess could be accompanied by a $\nu_\mu$ or $\overline{\nu}_\mu$ excess.
The question of whether this muon component can be observed is not straightforward.
In principle, such an excess could lead to a higher rate of muons in MiniBooNE, modifying the measured distribution of $\nu_\mu$CC events.
Because of the different kinematics of neutrinos and antineutrinos, the energy of a $\overline{\nu}_\mu$ could be mis-reconstructed as a higher-energy $\nu_\mu$.
In addition, resonant and coherent charged-pion production would also be modified.
By studying CC$\pi^+$ production, MiniBooNE used this method to constrain the wrong-sign $\nu_\mu$ CCQE events in antineutrino mode~\cite{MiniBooNE:2020pnu}.
In practice, however, a very low-energy flux excess like that in \cref{fig:unfolded_templates} (top) would happen in a region close to the kinematic threshold of muon and pion production.
This could exacerbate the excess in $\overline{\nu}_e$CC events, where threshold effects are not important.
Elastic processes like neutrino- and antineutrino-electron scattering would be impacted, but this component makes up less than $2\%$ of the total number of electron-like events observed by MiniBooNE~\cite{MiniBooNE:2020pnu}.
Finally, we note that when the MicroBooNE Wire-Cell inclusive $1eX$ sample is divided into $0pX\pi$, and $NpX\pi$ events, a small excess of $\nu_\mu$CC $0pX\pi$ events is observed in the energy region of $0.3 - 1.1$~GeV~\cite{MicroBooNE:2021nxr}.
Because $\nu_\mu$CC $NpX\pi$ events are in good agreement with the Monte Carlo, this effect, if it grows in significance, could be explained due to an excess of $\overline{\nu}_\mu$ in the BNB.

To conclude this section, we have found that the full $\overline\nu_e$ excess shown \cref{fig:flux_excess} would require a significant deviation from the BNB model presented in Ref.~\cite{MiniBooNE:2008hfu}.
We remind the reader that our discussion focused on neutrino mode only.
Since the origin of the excess in neutrino and antineutrino modes are both unknown, the neutrino-to-antineutrino ratio of the excess in each mode is, in principle, not necessarily the same.
For that reason, we do not derive constraints on the excess in antineutrino mode using MicroBooNE data, as the latter was obtained with the beam in neutrino mode.
We leave a detailed study of possible anomalous sources of wrong-sign neutrinos in each mode to future work.

\subsection{New physics}

We now comment on a few possibilities for a Beyond-the-Standard-Model (BSM) origin of the $\overline{\nu}_e$ excess in MiniBooNE.
Neutrino-antineutrino oscillations $\nu_\mu \to \overline{\nu}_e$ can convert left-handed neutrinos into right-handed antineutrinos~\cite{Pontecorvo:1967fh}.
However, they require a chirality flip and are usually too small to be observable due to the smallness of $m_\nu^2/E_\nu^2$.
Large neutrino magnetic moments in strong magnetic fields can also induce such oscillations through spin-flip precession~\cite{Cisneros:1970nq,Voloshin:1986ty,Okun:1986uf} but are not relevant for short-baseline experiments.
In general, $\nu \to \overline{\nu}$ oscillations are constrained experimentally by solar neutrino experiments~\cite{Borexino:2019wln,Super-Kamiokande:2020frs,KamLAND:2021gvi} and by direct searches~\cite{Cooper-Sarkar:1981bam}.

A $\overline{\nu}_e$ excess could also stem from exotic pion decays at the target.
For instance, the lepton-flavor- and lepton-number-violating branching ratio for the pion, $\pi^+\to\mu^+ \overline{\nu}_e$, could produce $\overline{\nu}_e$ with the same energy spectrum as the neutrinos from $\pi^+\to \mu^+ \nu_\mu$.
The best experimental limits on this decay come from the BEBC detector, which sat in the wide-band high-energy neutrino beam at CERN~\cite{Cooper-Sarkar:1981bam}.
A dedicated search finds $\mathcal{B}(\pi^+\to \mu^+ \overline{\nu}_e) < 0.15\%$ and $\mathcal{B}(\pi^+\to \mu^+ \nu_e) < 0.8\%$ at $90\%$~CL.
Precision tests of lepton flavor universality are also sensitive to this decay channel.
Comparing the SM prediction~\cite{Cirigliano:2007xi} with the experimental measurements~\cite{Bryman:1985bv,Czapek:1993kc,Britton:1993cj,PiENu:2015seu} of $R_{e/\mu} = \Gamma(\pi^+\to e^+ \nu)/\Gamma(\pi^+\to \mu^+ \nu)$, we find
\begin{equation}
    \frac{\Gamma(\pi^+\to \mu^+ \overline{\nu}_{e})}{\Gamma(\pi^+\to \mu^+ \nu_\mu)} = (0.20 \pm 0.19)\%,
\end{equation}
providing stronger limits on the branching ratio, $\mathcal{B}(\pi^+\to \mu^+ \overline{\nu}_e) < 0.50\%$ at $90\%$~C.L.
We note that this branching ratio can also be constrained by the LSND experiment using low-energy IBD events; in the energy of interest, $E_{e^+} < E_{\overline{\nu}_e}^{\pi\text{DAR}} \sim 30$~MeV, good agreement is found between data and Monte Carlo.
We leave a careful evaluation of this limit for future work.

Light particles produced in the beam could be another source of antineutrinos.
Decaying light sterile neutrinos can produce antineutrinos in two scenarios.
If the sterile neutrino is a Majorana particle, antineutrinos can be produced in the decay $\nu_4 \to \overline{\nu} \phi$~\cite{Palomares-Ruiz:2005zbh,Moss:2017pur,Dentler:2019dhz,deGouvea:2019qre}, where $\phi$ is a neutrinophilic light scalar particle.
If the sterile neutrino is a Dirac particle, then the subsequent decays of the scalar particle, $\nu_4 \to \nu (\phi \to \overline{\nu}\nu)$, can produce antineutrinos at a fraction of $1:2$~\cite{Moss:2017pur,Dentler:2019dhz}.
Finally, lepton-number-charged scalars, $\phi_2$, can lead to antineutrinos in the decay of $\nu_4 \to \overline{\nu}\phi_2$.
In detection, the emission of $\phi_2$ can lead to what is effectively an off-shell antineutrino scattering process, $\nu_\mu p^+\to \phi_2 (\overline{\nu}_e)^* p^+ \to \phi_2 e^+ n$.
These explanations of MiniBooNE, however, are excluded by solar antineutrino searches~\cite{Hostert:2020oui} and meson decays~\cite{Berryman:2018ogk}.
We note that a decaying-sterile neutrino has been recently searched for by the IceCube Neutrino Observatory, finding a preference for decay~\cite{IceCubeCollaboration:2022tso}.
Though the analysis uses invisible decay, it is largely insensitive to the decay being visible or invisible due to the steeply falling spectra as discussed in~\cite{Moss:2017pur}; see~\cite{Hardin:2022muu} for a recent discussion of the latter scenario.

\subsection{Future prospects}

This study focuses specifically on the possibility of a $\overline{\nu}_e$ excess in the BNB; thus, it is relevant to consider whether future experiments along the BNB will be sensitive to such an excess.
MicroBooNE is part of the short baseline neutrino program at Fermilab, which includes the upcoming ICARUS and SBND experiments~\cite{MicroBooNE:2015bmn}.
ICARUS and SBND also use LArTPC detectors, so they too will suffer from the $\overline{\nu}_e$-Ar cross section suppression at low energies.
Of the SBN experiments, SBND in particular is the most optimistic setup to search for a $\overline{\nu}_e$ excess.
This is because it will benefit from a ten-fold enhancement in event rate compared to MicroBooNE, as it is situated closer to the BNB target~\cite{MicroBooNE:2015bmn}.
Even so, assuming SBND reconstruction efficiency is similar to MicroBooNE, a factor of $\sim 10$ enhancement in the backgrounds and excess templates shown in the right panels of \cref{fig:templates} will not significantly improve the sensitivity to \emph{Scenario 3}, unless a substantial reduction of the backgrounds can be achieved.

In view of the challenges in detecting antineutrinos in LArTPC detectors along the BNB, we turn to a different kind of detector for this measurement: the Accelerator Neutrino Neutron Interaction Experiment (ANNIE).
ANNIE is a 26-ton water Cherenkov detector located at $100$~m from the BNB target~\cite{ANNIE:2015inw,ANNIE:2017nng}.
The water volume is followed by a muon range detector to allow the detection of muon neutrino and antineutrino interactions.
One of the primary goals of ANNIE is to measure the neutron multiplicity in CCQE interactions.
For that, the detector is doped with gadolinium, so that neutrons produced in neutrino events can be detected via delayed $\sim8$~MeV photons emmited in neutron-gadolinium capture.

While ANNIE has so far focused on muon events, the detector can also measure inverse-beta-decay events, $\overline{\nu}_e p^+ \to e^+ n$.
The signature is a single, low-energy positron Cherenkov ring with a delayed neutron capture.
The photo coverage of ANNIE phase-II, in particular, provides the right environment for this measurement and can be used on a search for a $\overline{\nu}_e$ interpretation of the LEE at the current location of the detector.
A detailed background study is needed to estimate ANNIE's sensitivity to the excess flux in \cref{fig:flux_excess}.
Nevertheless, the addition of water-based liquid scintillator in the detector volume, ANNIE phase-III, would be a clear improvement to mitigate backgrounds~\cite{ANNIE:2017nng}.
Finally, because of the sheer magnitude of the excess of $\overline{\nu}_e$ required by the MiniBooNE LEE and the lack of information on its size below $E_\nu < 200$~MeV, ANNIE can start to probe $\overline{\nu}_e$-based explanations of the MiniBooNE LEE even if it is unable to detect the intrinsic $\overline{\nu}_e$ flux.

\section{Conclusion}
\label{sec:conclusions}

The MicroBooNE experiment has not found any evidence for an electron-neutrino interpretation of the MiniBooNE low-energy excess (LEE).
In this article, we show that this fact can be reconciled with the MiniBooNE observation if the LEE is caused by electron-antineutrinos instead.
This is due to three main reasons,
i) two out of the three MicroBooNE analyses have focused on single proton final states, and these are rarely produced in antineutrino-nucleus scattering,
ii) the energy of the initial $\overline{\nu}_e$ is substantially under-reconstructed in MicroBooNE due to the invisible final state neutron,
iii) the antineutrino cross sections per-nucleon on $^{40}$Ar are suppressed with respect to those in CH$_2$, the nuclear targets in MiniBooNE.
The differences in total cross section are due to the difference in the proton-to-neutron ratio ($4:3$ at MiniBooNE compared to $9:10$ at MicroBooNE), but, more importantly, due to the larger size of the argon nucleus and its non-isoscalar nature.
Contrary to carbon, the proton separation energy in argon is larger than that of neutrons, requiring a larger energy transfer in $\overline \nu$ CCQE scattering.
As shown in the bottom panel of \cref{fig:cross_sections}, this threshold effect is particularly significant in the energy region of the MiniBooNE LEE.

To quantify the impact of a $\overline{\nu}_e$-interpretation of the LEE on the latest MicroBooNE results, we followed Ref.~\cite{Arguelles:2021meu} and reproduced the results of the MicroBooNE Wire-Cell template analysis~\cite{MicroBooNE:2021nxr}.
Because antineutrinos do not produce protons, we focused on the inclusive Wire-Cell sample, which does not require a proton connected to the neutrino interaction vertex.
We then estimated new detector response matrices under the assumption of electron-antineutrino charged-current scattering and proceeded to unfold the MiniBooNE LEE into a $\overline{\nu}_e$LEE template, shown in the top panel of \cref{fig:unfolded_templates}.
We checked that the unfolded template reproduces the MiniBooNE LEE once folded back into MiniBooNE with our response matrix, as shown in the bottom panel of \cref{fig:unfolded_templates}.

We showed that if the antineutrino-to-neutrino ratio of the LEE event rate is 100\% ($\f = 1$), then MicroBooNE's sensitivity is significantly reduced to less than $2\sigma$ (assuming Wilks' theorem), as much fewer LEE events are expected in the detector.
If the number of antineutrino-induced LEE events is $50\%$, then MicroBooNE's sensitivity to the nominal template is reduced to less than $3\sigma$ C.L. for the nominal LEE template.
As pointed out in Ref.~\cite{Arguelles:2021meu}, due to the large background systematic uncertainties in MiniBooNE, choosing different templates with an excellent fit to the LEE, $p_{\rm val}^{\rm LEE} > 90\%$, can have a significant impact on MicroBooNE's sensitivity.
Using the best and worst-case template choices from Ref.~\cite{Arguelles:2021meu}, if we take the MiniBooNE excess at face value and require $x=1.0$ the MicroBooNE data constrain the antineutrino-to-neutrino fraction of the LEE event rate to be at least $\fLowerBest$ for template T1 (best-case scenario) and $\fLowerWorst$ for template T2 (worst-case scenario) at the $2\sigma$~C.L., assuming Wilks' theorem for 1~d.o.f.

The Deep-Learning analysis focused specifically on the $1e1p$ event topology, maximizing its sensitivity to electron-neutrino CCQE events.
While this analysis had the largest purity, the requirement of a final state proton makes it insensitive to antineutrinos, which can only produce a proton through nuclear final state interactions.
Finally, the Pandora analysis focused on pionless topologies, separating them into $1e0\pi Np$ and $1e0\pi0p$.
While the latter does not require a proton in the final state, it is also the least sensitive and less pure of the analyses.
It is also the only one that observes a small excess of events.
We can then conclude that the choice of event topologies makes the Pandora and Deep-Learning analyses insensitive to an excess of $\overline{\nu}_e$.

The other LArTPC detectors along the BNB, SBND and ICARUS, will face the same issues as MicroBooNE when testing a $\overline{\nu}_e$ explanation of the MiniBooNE LEE, due to the $\overline{\nu}_e$-Ar cross section suppression at low energy.
In principle, the near detector of the SBN program, SBND, will observe more antineutrino events but on top of a higher overall event rate.
In light of this, we have pointed out a different possibility to directly search for $\overline{\nu}_e$ in the BNB using the ANNIE detector.
The phase-II of ANNIE is particularly well-suited for the study of low-energy inverse-beta-decay.
The scattering of antineutrinos on free protons inside the water-based Cherenkov detector produces a prompt positron signal followed by a delayed capture of neutrons on gadolinium.
A detailed study of the backgrounds is needed to assess the final sensitivity of the experiment.
Further improvements would be possible with phase-III, where the separation of scintillation and Cherenkov light could be achieved with a water-based liquid scintillator volume.
A dedicated analysis at ANNIE can shed new light on SM as well as BSM explanations of the MiniBooNE LEE, targeting the BNB flux at energies of the LEE and below.

\acknowledgements
We want to thank Mayly~C.~Sanchez and Michael~Wurm for interesting discussions on the ANNIE experiment.
The research at the Perimeter Institute is supported in part by the Government of Canada through NSERC and by the Province of Ontario through the Ministry of Economic Development, Job Creation and Trade, MEDT.
CAA are supported by the Faculty of Arts and Sciences of Harvard University.
NWK is supported by the NSF Graduate Research Fellowship under Grant No. 1745302.
JMC and MHS are supported by the NSF.

\bibliographystyle{apsrev4-1}
\bibliography{main}

\begin{thebibliography}{48}%
\makeatletter
\providecommand \@ifxundefined [1]{%
 \@ifx{#1\undefined}
}%
\providecommand \@ifnum [1]{%
 \ifnum #1\expandafter \@firstoftwo
 \else \expandafter \@secondoftwo
 \fi
}%
\providecommand \@ifx [1]{%
 \ifx #1\expandafter \@firstoftwo
 \else \expandafter \@secondoftwo
 \fi
}%
\providecommand \natexlab [1]{#1}%
\providecommand \enquote  [1]{``#1''}%
\providecommand \bibnamefont  [1]{#1}%
\providecommand \bibfnamefont [1]{#1}%
\providecommand \citenamefont [1]{#1}%
\providecommand \href@noop [0]{\@secondoftwo}%
\providecommand \href [0]{\begingroup \@sanitize@url \@href}%
\providecommand \@href[1]{\@@startlink{#1}\@@href}%
\providecommand \@@href[1]{\endgroup#1\@@endlink}%
\providecommand \@sanitize@url [0]{\catcode `\\12\catcode `\$12\catcode
  `\&12\catcode `\#12\catcode `\^12\catcode `\_12\catcode `\%12\relax}%
\providecommand \@@startlink[1]{}%
\providecommand \@@endlink[0]{}%
\providecommand \url  [0]{\begingroup\@sanitize@url \@url }%
\providecommand \@url [1]{\endgroup\@href {#1}{\urlprefix }}%
\providecommand \urlprefix  [0]{URL }%
\providecommand \Eprint [0]{\href }%
\providecommand \doibase [0]{http://dx.doi.org/}%
\providecommand \selectlanguage [0]{\@gobble}%
\providecommand \bibinfo  [0]{\@secondoftwo}%
\providecommand \bibfield  [0]{\@secondoftwo}%
\providecommand \translation [1]{[#1]}%
\providecommand \BibitemOpen [0]{}%
\providecommand \bibitemStop [0]{}%
\providecommand \bibitemNoStop [0]{.\EOS\space}%
\providecommand \EOS [0]{\spacefactor3000\relax}%
\providecommand \BibitemShut  [1]{\csname bibitem#1\endcsname}%
\let\auto@bib@innerbib\@empty
\bibitem [{\citenamefont {Aguilar-Arevalo}\ \emph {et~al.}(2009)\citenamefont
  {Aguilar-Arevalo} \emph {et~al.}}]{MiniBooNE:2008hfu}%
  \BibitemOpen
  \bibfield  {author} {\bibinfo {author} {\bibfnamefont {A.~A.}\ \bibnamefont
  {Aguilar-Arevalo}} \emph {et~al.} (\bibinfo {collaboration} {MiniBooNE}),\
  }\href {\doibase 10.1103/PhysRevD.79.072002} {\bibfield  {journal} {\bibinfo
  {journal} {Phys. Rev. D}\ }\textbf {\bibinfo {volume} {79}},\ \bibinfo
  {pages} {072002} (\bibinfo {year} {2009})},\ \Eprint
  {http://arxiv.org/abs/0806.1449} {arXiv:0806.1449 [hep-ex]} \BibitemShut
  {NoStop}%
\bibitem [{\citenamefont {Aguilar-Arevalo}\ \emph
  {et~al.}(2021{\natexlab{a}})\citenamefont {Aguilar-Arevalo} \emph
  {et~al.}}]{MiniBooNE:2020pnu}%
  \BibitemOpen
  \bibfield  {author} {\bibinfo {author} {\bibfnamefont {A.~A.}\ \bibnamefont
  {Aguilar-Arevalo}} \emph {et~al.} (\bibinfo {collaboration} {MiniBooNE}),\
  }\href {\doibase 10.1103/PhysRevD.103.052002} {\bibfield  {journal} {\bibinfo
   {journal} {Phys. Rev. D}\ }\textbf {\bibinfo {volume} {103}},\ \bibinfo
  {pages} {052002} (\bibinfo {year} {2021}{\natexlab{a}})},\ \Eprint
  {http://arxiv.org/abs/2006.16883} {arXiv:2006.16883 [hep-ex]} \BibitemShut
  {NoStop}%
\bibitem [{\citenamefont {Brdar}\ and\ \citenamefont
  {Kopp}(2022)}]{Brdar:2021ysi}%
  \BibitemOpen
  \bibfield  {author} {\bibinfo {author} {\bibfnamefont {V.}~\bibnamefont
  {Brdar}}\ and\ \bibinfo {author} {\bibfnamefont {J.}~\bibnamefont {Kopp}},\
  }\href {\doibase 10.1103/PhysRevD.105.115024} {\bibfield  {journal} {\bibinfo
   {journal} {Phys. Rev. D}\ }\textbf {\bibinfo {volume} {105}},\ \bibinfo
  {pages} {115024} (\bibinfo {year} {2022})},\ \Eprint
  {http://arxiv.org/abs/2109.08157} {arXiv:2109.08157 [hep-ph]} \BibitemShut
  {NoStop}%
\bibitem [{\citenamefont {Kelly}\ and\ \citenamefont
  {Kopp}(2022)}]{Kelly:2022uaa}%
  \BibitemOpen
  \bibfield  {author} {\bibinfo {author} {\bibfnamefont {K.~J.}\ \bibnamefont
  {Kelly}}\ and\ \bibinfo {author} {\bibfnamefont {J.}~\bibnamefont {Kopp}},\
  }\href@noop {} {\  (\bibinfo {year} {2022})},\ \Eprint
  {http://arxiv.org/abs/2210.08021} {arXiv:2210.08021 [hep-ph]} \BibitemShut
  {NoStop}%
\bibitem [{\citenamefont {Acero}\ \emph {et~al.}(2022)\citenamefont {Acero}
  \emph {et~al.}}]{Acero:2022wqg}%
  \BibitemOpen
  \bibfield  {author} {\bibinfo {author} {\bibfnamefont {M.~A.}\ \bibnamefont
  {Acero}} \emph {et~al.},\ }\href@noop {} {\  (\bibinfo {year} {2022})},\
  \Eprint {http://arxiv.org/abs/2203.07323} {arXiv:2203.07323 [hep-ex]}
  \BibitemShut {NoStop}%
\bibitem [{\citenamefont {Aguilar-Arevalo}\ \emph {et~al.}(2001)\citenamefont
  {Aguilar-Arevalo} \emph {et~al.}}]{LSND:2001aii}%
  \BibitemOpen
  \bibfield  {author} {\bibinfo {author} {\bibfnamefont {A.}~\bibnamefont
  {Aguilar-Arevalo}} \emph {et~al.} (\bibinfo {collaboration} {LSND}),\ }\href
  {\doibase 10.1103/PhysRevD.64.112007} {\bibfield  {journal} {\bibinfo
  {journal} {Phys. Rev. D}\ }\textbf {\bibinfo {volume} {64}},\ \bibinfo
  {pages} {112007} (\bibinfo {year} {2001})},\ \Eprint
  {http://arxiv.org/abs/hep-ex/0104049} {arXiv:hep-ex/0104049} \BibitemShut
  {NoStop}%
\bibitem [{\citenamefont {Chen}\ \emph {et~al.}(2007)\citenamefont {Chen} \emph
  {et~al.}}]{MicroBooNE:2007ivj}%
  \BibitemOpen
  \bibfield  {author} {\bibinfo {author} {\bibfnamefont {H.}~\bibnamefont
  {Chen}} \emph {et~al.} (\bibinfo {collaboration} {MicroBooNE}),\ }\href@noop
  {} {\  (\bibinfo {year} {2007})}\BibitemShut {NoStop}%
\bibitem [{\citenamefont {Abratenko}\ \emph
  {et~al.}(2021{\natexlab{a}})\citenamefont {Abratenko} \emph
  {et~al.}}]{MicroBooNE:2020sar}%
  \BibitemOpen
  \bibfield  {author} {\bibinfo {author} {\bibfnamefont {P.}~\bibnamefont
  {Abratenko}} \emph {et~al.} (\bibinfo {collaboration} {MicroBooNE}),\ }\href
  {\doibase 10.1088/1748-0221/16/02/P02017} {\bibfield  {journal} {\bibinfo
  {journal} {JINST}\ }\textbf {\bibinfo {volume} {16}},\ \bibinfo {pages}
  {P02017} (\bibinfo {year} {2021}{\natexlab{a}})},\ \Eprint
  {http://arxiv.org/abs/2002.09375} {arXiv:2002.09375 [physics.ins-det]}
  \BibitemShut {NoStop}%
\bibitem [{\citenamefont {Abratenko}\ \emph
  {et~al.}(2022{\natexlab{a}})\citenamefont {Abratenko} \emph
  {et~al.}}]{MicroBooNE:2021rmx}%
  \BibitemOpen
  \bibfield  {author} {\bibinfo {author} {\bibfnamefont {P.}~\bibnamefont
  {Abratenko}} \emph {et~al.} (\bibinfo {collaboration} {MicroBooNE}),\ }\href
  {\doibase 10.1103/PhysRevLett.128.241801} {\bibfield  {journal} {\bibinfo
  {journal} {Phys. Rev. Lett.}\ }\textbf {\bibinfo {volume} {128}},\ \bibinfo
  {pages} {241801} (\bibinfo {year} {2022}{\natexlab{a}})},\ \Eprint
  {http://arxiv.org/abs/2110.14054} {arXiv:2110.14054 [hep-ex]} \BibitemShut
  {NoStop}%
\bibitem [{\citenamefont {Abratenko}\ \emph {et~al.}(2018)\citenamefont
  {Abratenko} \emph {et~al.}}]{MicroBooNE:2018vnm}%
  \BibitemOpen
  \bibfield  {author} {\bibinfo {author} {\bibfnamefont {P.}~\bibnamefont
  {Abratenko}} \emph {et~al.} (\bibinfo {collaboration} {MicroBooNE}),\ }\href
  {\doibase 10.2172/1573217} {\  (\bibinfo {year} {2018}),\
  10.2172/1573217}\BibitemShut {NoStop}%
\bibitem [{\citenamefont {D'Agostini}(1995)}]{DAgostini:1994fjx}%
  \BibitemOpen
  \bibfield  {author} {\bibinfo {author} {\bibfnamefont {G.}~\bibnamefont
  {D'Agostini}},\ }\href {\doibase 10.1016/0168-9002(95)00274-X} {\bibfield
  {journal} {\bibinfo  {journal} {Nucl. Instrum. Meth. A}\ }\textbf {\bibinfo
  {volume} {362}},\ \bibinfo {pages} {487} (\bibinfo {year}
  {1995})}\BibitemShut {NoStop}%
\bibitem [{\citenamefont {Arg\"uelles}\ \emph {et~al.}(2022)\citenamefont
  {Arg\"uelles}, \citenamefont {Esteban}, \citenamefont {Hostert},
  \citenamefont {Kelly}, \citenamefont {Kopp}, \citenamefont {Machado},
  \citenamefont {Martinez-Soler},\ and\ \citenamefont
  {Perez-Gonzalez}}]{Arguelles:2021meu}%
  \BibitemOpen
  \bibfield  {author} {\bibinfo {author} {\bibfnamefont {C.~A.}\ \bibnamefont
  {Arg\"uelles}}, \bibinfo {author} {\bibfnamefont {I.}~\bibnamefont
  {Esteban}}, \bibinfo {author} {\bibfnamefont {M.}~\bibnamefont {Hostert}},
  \bibinfo {author} {\bibfnamefont {K.~J.}\ \bibnamefont {Kelly}}, \bibinfo
  {author} {\bibfnamefont {J.}~\bibnamefont {Kopp}}, \bibinfo {author}
  {\bibfnamefont {P.~A.~N.}\ \bibnamefont {Machado}}, \bibinfo {author}
  {\bibfnamefont {I.}~\bibnamefont {Martinez-Soler}}, \ and\ \bibinfo {author}
  {\bibfnamefont {Y.~F.}\ \bibnamefont {Perez-Gonzalez}},\ }\href {\doibase
  10.1103/PhysRevLett.128.241802} {\bibfield  {journal} {\bibinfo  {journal}
  {Phys. Rev. Lett.}\ }\textbf {\bibinfo {volume} {128}},\ \bibinfo {pages}
  {241802} (\bibinfo {year} {2022})},\ \Eprint
  {http://arxiv.org/abs/2111.10359} {arXiv:2111.10359 [hep-ph]} \BibitemShut
  {NoStop}%
\bibitem [{\citenamefont {Abratenko}\ \emph
  {et~al.}(2022{\natexlab{b}})\citenamefont {Abratenko} \emph
  {et~al.}}]{MicroBooNE:2021bcu}%
  \BibitemOpen
  \bibfield  {author} {\bibinfo {author} {\bibfnamefont {P.}~\bibnamefont
  {Abratenko}} \emph {et~al.} (\bibinfo {collaboration} {MicroBooNE}),\ }\href
  {\doibase 10.1103/PhysRevD.105.112003} {\bibfield  {journal} {\bibinfo
  {journal} {Phys. Rev. D}\ }\textbf {\bibinfo {volume} {105}},\ \bibinfo
  {pages} {112003} (\bibinfo {year} {2022}{\natexlab{b}})},\ \Eprint
  {http://arxiv.org/abs/2110.14080} {arXiv:2110.14080 [hep-ex]} \BibitemShut
  {NoStop}%
\bibitem [{\citenamefont {Abratenko}\ \emph
  {et~al.}(2022{\natexlab{c}})\citenamefont {Abratenko} \emph
  {et~al.}}]{MicroBooNE:2021nxr}%
  \BibitemOpen
  \bibfield  {author} {\bibinfo {author} {\bibfnamefont {P.}~\bibnamefont
  {Abratenko}} \emph {et~al.} (\bibinfo {collaboration} {MicroBooNE}),\ }\href
  {\doibase 10.1103/PhysRevD.105.112005} {\bibfield  {journal} {\bibinfo
  {journal} {Phys. Rev. D}\ }\textbf {\bibinfo {volume} {105}},\ \bibinfo
  {pages} {112005} (\bibinfo {year} {2022}{\natexlab{c}})},\ \Eprint
  {http://arxiv.org/abs/2110.13978} {arXiv:2110.13978 [hep-ex]} \BibitemShut
  {NoStop}%
\bibitem [{\citenamefont {Abratenko}\ \emph
  {et~al.}(2021{\natexlab{b}})\citenamefont {Abratenko} \emph
  {et~al.}}]{MicroBooNE:2020yze}%
  \BibitemOpen
  \bibfield  {author} {\bibinfo {author} {\bibfnamefont {P.}~\bibnamefont
  {Abratenko}} \emph {et~al.} (\bibinfo {collaboration} {MicroBooNE}),\ }\href
  {\doibase 10.1103/PhysRevD.103.052012} {\bibfield  {journal} {\bibinfo
  {journal} {Phys. Rev. D}\ }\textbf {\bibinfo {volume} {103}},\ \bibinfo
  {pages} {052012} (\bibinfo {year} {2021}{\natexlab{b}})},\ \Eprint
  {http://arxiv.org/abs/2012.08513} {arXiv:2012.08513 [physics.ins-det]}
  \BibitemShut {NoStop}%
\bibitem [{\citenamefont {Abratenko}\ \emph
  {et~al.}(2021{\natexlab{c}})\citenamefont {Abratenko} \emph
  {et~al.}}]{MicroBooNE:2020hho}%
  \BibitemOpen
  \bibfield  {author} {\bibinfo {author} {\bibfnamefont {P.}~\bibnamefont
  {Abratenko}} \emph {et~al.} (\bibinfo {collaboration} {MicroBooNE}),\ }\href
  {\doibase 10.1103/PhysRevD.103.092003} {\bibfield  {journal} {\bibinfo
  {journal} {Phys. Rev. D}\ }\textbf {\bibinfo {volume} {103}},\ \bibinfo
  {pages} {092003} (\bibinfo {year} {2021}{\natexlab{c}})},\ \Eprint
  {http://arxiv.org/abs/2010.08653} {arXiv:2010.08653 [hep-ex]} \BibitemShut
  {NoStop}%
\bibitem [{\citenamefont {Abratenko}\ \emph
  {et~al.}(2022{\natexlab{d}})\citenamefont {Abratenko} \emph
  {et~al.}}]{MicroBooNE:2021ojx}%
  \BibitemOpen
  \bibfield  {author} {\bibinfo {author} {\bibfnamefont {P.}~\bibnamefont
  {Abratenko}} \emph {et~al.} (\bibinfo {collaboration} {MicroBooNE}),\ }\href
  {\doibase 10.1088/1748-0221/17/01/P01037} {\bibfield  {journal} {\bibinfo
  {journal} {JINST}\ }\textbf {\bibinfo {volume} {17}},\ \bibinfo {pages}
  {P01037} (\bibinfo {year} {2022}{\natexlab{d}})},\ \Eprint
  {http://arxiv.org/abs/2110.13961} {arXiv:2110.13961 [physics.ins-det]}
  \BibitemShut {NoStop}%
\bibitem [{\citenamefont {Andreopoulos}\ \emph {et~al.}(2015)\citenamefont
  {Andreopoulos}, \citenamefont {Barry}, \citenamefont {Dytman}, \citenamefont
  {Gallagher}, \citenamefont {Golan}, \citenamefont {Hatcher}, \citenamefont
  {Perdue},\ and\ \citenamefont {Yarba}}]{Andreopoulos:2015wxa}%
  \BibitemOpen
  \bibfield  {author} {\bibinfo {author} {\bibfnamefont {C.}~\bibnamefont
  {Andreopoulos}}, \bibinfo {author} {\bibfnamefont {C.}~\bibnamefont {Barry}},
  \bibinfo {author} {\bibfnamefont {S.}~\bibnamefont {Dytman}}, \bibinfo
  {author} {\bibfnamefont {H.}~\bibnamefont {Gallagher}}, \bibinfo {author}
  {\bibfnamefont {T.}~\bibnamefont {Golan}}, \bibinfo {author} {\bibfnamefont
  {R.}~\bibnamefont {Hatcher}}, \bibinfo {author} {\bibfnamefont
  {G.}~\bibnamefont {Perdue}}, \ and\ \bibinfo {author} {\bibfnamefont
  {J.}~\bibnamefont {Yarba}},\ }\href@noop {} {\  (\bibinfo {year} {2015})},\
  \Eprint {http://arxiv.org/abs/1510.05494} {arXiv:1510.05494 [hep-ph]}
  \BibitemShut {NoStop}%
\bibitem [{\citenamefont {Alvarez-Ruso}\ \emph {et~al.}(2021)\citenamefont
  {Alvarez-Ruso} \emph {et~al.}}]{GENIE:2021npt}%
  \BibitemOpen
  \bibfield  {author} {\bibinfo {author} {\bibfnamefont {L.}~\bibnamefont
  {Alvarez-Ruso}} \emph {et~al.} (\bibinfo {collaboration} {GENIE}),\ }\href
  {\doibase 10.1140/epjs/s11734-021-00295-7} {\bibfield  {journal} {\bibinfo
  {journal} {Eur. Phys. J. ST}\ }\textbf {\bibinfo {volume} {230}},\ \bibinfo
  {pages} {4449} (\bibinfo {year} {2021})},\ \Eprint
  {http://arxiv.org/abs/2106.09381} {arXiv:2106.09381 [hep-ph]} \BibitemShut
  {NoStop}%
\bibitem [{\citenamefont {Abratenko}\ \emph
  {et~al.}(2022{\natexlab{e}})\citenamefont {Abratenko} \emph
  {et~al.}}]{MicroBooNE:2021ccs}%
  \BibitemOpen
  \bibfield  {author} {\bibinfo {author} {\bibfnamefont {P.}~\bibnamefont
  {Abratenko}} \emph {et~al.} (\bibinfo {collaboration} {MicroBooNE}),\ }\href
  {\doibase 10.1103/PhysRevD.105.072001} {\bibfield  {journal} {\bibinfo
  {journal} {Phys. Rev. D}\ }\textbf {\bibinfo {volume} {105}},\ \bibinfo
  {pages} {072001} (\bibinfo {year} {2022}{\natexlab{e}})},\ \Eprint
  {http://arxiv.org/abs/2110.14028} {arXiv:2110.14028 [hep-ex]} \BibitemShut
  {NoStop}%
\bibitem [{\citenamefont {Pavlovic}\ \emph {et~al.}(2012)\citenamefont
  {Pavlovic}, \citenamefont {Van~de Water},\ and\ \citenamefont
  {Zeller}}]{MiniBooNE_eff}%
  \BibitemOpen
  \bibfield  {author} {\bibinfo {author} {\bibfnamefont {Z.}~\bibnamefont
  {Pavlovic}}, \bibinfo {author} {\bibfnamefont {R.}~\bibnamefont {Van~de
  Water}}, \ and\ \bibinfo {author} {\bibfnamefont {S.}~\bibnamefont
  {Zeller}},\ }\href@noop {} {\emph {\bibinfo {title} {MiniBooNE Gamma-Ray and
  Electron Efficiencies}}},\ \bibinfo {type} {Tech. Rep.}\ (\bibinfo {year}
  {2012})\BibitemShut {NoStop}%
\bibitem [{\citenamefont {Aguilar-Arevalo}\ \emph {et~al.}(2018)\citenamefont
  {Aguilar-Arevalo} \emph {et~al.}}]{MiniBooNE:2018esg}%
  \BibitemOpen
  \bibfield  {author} {\bibinfo {author} {\bibfnamefont {A.~A.}\ \bibnamefont
  {Aguilar-Arevalo}} \emph {et~al.} (\bibinfo {collaboration} {MiniBooNE}),\
  }\href {\doibase 10.1103/PhysRevLett.121.221801} {\bibfield  {journal}
  {\bibinfo  {journal} {Phys. Rev. Lett.}\ }\textbf {\bibinfo {volume} {121}},\
  \bibinfo {pages} {221801} (\bibinfo {year} {2018})},\ \Eprint
  {http://arxiv.org/abs/1805.12028} {arXiv:1805.12028 [hep-ex]} \BibitemShut
  {NoStop}%
\bibitem [{\citenamefont {Wilks}(1938)}]{Wilks:1938dza}%
  \BibitemOpen
  \bibfield  {author} {\bibinfo {author} {\bibfnamefont {S.~S.}\ \bibnamefont
  {Wilks}},\ }\href {\doibase 10.1214/aoms/1177732360} {\bibfield  {journal}
  {\bibinfo  {journal} {Annals Math. Statist.}\ }\textbf {\bibinfo {volume}
  {9}},\ \bibinfo {pages} {60} (\bibinfo {year} {1938})}\BibitemShut {NoStop}%
\bibitem [{\citenamefont {Pontecorvo}(1967)}]{Pontecorvo:1967fh}%
  \BibitemOpen
  \bibfield  {author} {\bibinfo {author} {\bibfnamefont {B.}~\bibnamefont
  {Pontecorvo}},\ }\href@noop {} {\bibfield  {journal} {\bibinfo  {journal}
  {Zh. Eksp. Teor. Fiz.}\ }\textbf {\bibinfo {volume} {53}},\ \bibinfo {pages}
  {1717} (\bibinfo {year} {1967})}\BibitemShut {NoStop}%
\bibitem [{\citenamefont {Cisneros}(1971)}]{Cisneros:1970nq}%
  \BibitemOpen
  \bibfield  {author} {\bibinfo {author} {\bibfnamefont {A.}~\bibnamefont
  {Cisneros}},\ }\href {\doibase 10.1007/BF00654607} {\bibfield  {journal}
  {\bibinfo  {journal} {Astrophys. Space Sci.}\ }\textbf {\bibinfo {volume}
  {10}},\ \bibinfo {pages} {87} (\bibinfo {year} {1971})}\BibitemShut {NoStop}%
\bibitem [{\citenamefont {Voloshin}\ and\ \citenamefont
  {Vysotsky}(1986)}]{Voloshin:1986ty}%
  \BibitemOpen
  \bibfield  {author} {\bibinfo {author} {\bibfnamefont {M.~B.}\ \bibnamefont
  {Voloshin}}\ and\ \bibinfo {author} {\bibfnamefont {M.~I.}\ \bibnamefont
  {Vysotsky}},\ }\href@noop {} {\bibfield  {journal} {\bibinfo  {journal} {Sov.
  J. Nucl. Phys.}\ }\textbf {\bibinfo {volume} {44}},\ \bibinfo {pages} {544}
  (\bibinfo {year} {1986})}\BibitemShut {NoStop}%
\bibitem [{\citenamefont {Okun}(1986)}]{Okun:1986uf}%
  \BibitemOpen
  \bibfield  {author} {\bibinfo {author} {\bibfnamefont {L.~B.}\ \bibnamefont
  {Okun}},\ }\href@noop {} {\bibfield  {journal} {\bibinfo  {journal} {Sov. J.
  Nucl. Phys.}\ }\textbf {\bibinfo {volume} {44}},\ \bibinfo {pages} {546}
  (\bibinfo {year} {1986})}\BibitemShut {NoStop}%
\bibitem [{\citenamefont {Agostini}\ \emph {et~al.}(2021)\citenamefont
  {Agostini} \emph {et~al.}}]{Borexino:2019wln}%
  \BibitemOpen
  \bibfield  {author} {\bibinfo {author} {\bibfnamefont {M.}~\bibnamefont
  {Agostini}} \emph {et~al.} (\bibinfo {collaboration} {Borexino}),\ }\href
  {\doibase 10.1016/j.astropartphys.2020.102509} {\bibfield  {journal}
  {\bibinfo  {journal} {Astropart. Phys.}\ }\textbf {\bibinfo {volume} {125}},\
  \bibinfo {pages} {102509} (\bibinfo {year} {2021})},\ \Eprint
  {http://arxiv.org/abs/1909.02422} {arXiv:1909.02422 [hep-ex]} \BibitemShut
  {NoStop}%
\bibitem [{\citenamefont {Abe}\ \emph {et~al.}(2022{\natexlab{a}})\citenamefont
  {Abe} \emph {et~al.}}]{Super-Kamiokande:2020frs}%
  \BibitemOpen
  \bibfield  {author} {\bibinfo {author} {\bibfnamefont {K.}~\bibnamefont
  {Abe}} \emph {et~al.} (\bibinfo {collaboration} {Super-Kamiokande}),\ }\href
  {\doibase 10.1016/j.astropartphys.2022.102702} {\bibfield  {journal}
  {\bibinfo  {journal} {Astropart. Phys.}\ }\textbf {\bibinfo {volume} {139}},\
  \bibinfo {pages} {102702} (\bibinfo {year} {2022}{\natexlab{a}})},\ \Eprint
  {http://arxiv.org/abs/2012.03807} {arXiv:2012.03807 [hep-ex]} \BibitemShut
  {NoStop}%
\bibitem [{\citenamefont {Abe}\ \emph {et~al.}(2022{\natexlab{b}})\citenamefont
  {Abe} \emph {et~al.}}]{KamLAND:2021gvi}%
  \BibitemOpen
  \bibfield  {author} {\bibinfo {author} {\bibfnamefont {S.}~\bibnamefont
  {Abe}} \emph {et~al.} (\bibinfo {collaboration} {KamLAND}),\ }\href {\doibase
  10.3847/1538-4357/ac32c1} {\bibfield  {journal} {\bibinfo  {journal}
  {Astrophys. J.}\ }\textbf {\bibinfo {volume} {925}},\ \bibinfo {pages} {14}
  (\bibinfo {year} {2022}{\natexlab{b}})},\ \Eprint
  {http://arxiv.org/abs/2108.08527} {arXiv:2108.08527 [astro-ph.HE]}
  \BibitemShut {NoStop}%
\bibitem [{\citenamefont {Cooper-Sarkar}\ \emph {et~al.}(1982)\citenamefont
  {Cooper-Sarkar}, \citenamefont {Guy}, \citenamefont {Michette}, \citenamefont
  {Tyndel},\ and\ \citenamefont {Venus}}]{Cooper-Sarkar:1981bam}%
  \BibitemOpen
  \bibfield  {author} {\bibinfo {author} {\bibfnamefont {A.~M.}\ \bibnamefont
  {Cooper-Sarkar}}, \bibinfo {author} {\bibfnamefont {J.~G.}\ \bibnamefont
  {Guy}}, \bibinfo {author} {\bibfnamefont {A.~G.}\ \bibnamefont {Michette}},
  \bibinfo {author} {\bibfnamefont {M.}~\bibnamefont {Tyndel}}, \ and\ \bibinfo
  {author} {\bibfnamefont {W.}~\bibnamefont {Venus}},\ }\href {\doibase
  10.1016/0370-2693(82)90914-5} {\bibfield  {journal} {\bibinfo  {journal}
  {Phys. Lett. B}\ }\textbf {\bibinfo {volume} {112}},\ \bibinfo {pages} {97}
  (\bibinfo {year} {1982})}\BibitemShut {NoStop}%
\bibitem [{\citenamefont {Cirigliano}\ and\ \citenamefont
  {Rosell}(2007)}]{Cirigliano:2007xi}%
  \BibitemOpen
  \bibfield  {author} {\bibinfo {author} {\bibfnamefont {V.}~\bibnamefont
  {Cirigliano}}\ and\ \bibinfo {author} {\bibfnamefont {I.}~\bibnamefont
  {Rosell}},\ }\href {\doibase 10.1103/PhysRevLett.99.231801} {\bibfield
  {journal} {\bibinfo  {journal} {Phys. Rev. Lett.}\ }\textbf {\bibinfo
  {volume} {99}},\ \bibinfo {pages} {231801} (\bibinfo {year} {2007})},\
  \Eprint {http://arxiv.org/abs/0707.3439} {arXiv:0707.3439 [hep-ph]}
  \BibitemShut {NoStop}%
\bibitem [{\citenamefont {Bryman}\ \emph {et~al.}(1986)\citenamefont {Bryman},
  \citenamefont {Dixit}, \citenamefont {Dubois}, \citenamefont {Macdonald},
  \citenamefont {Numao}, \citenamefont {Olaniyi}, \citenamefont {Olin},\ and\
  \citenamefont {Poutissou}}]{Bryman:1985bv}%
  \BibitemOpen
  \bibfield  {author} {\bibinfo {author} {\bibfnamefont {D.~A.}\ \bibnamefont
  {Bryman}}, \bibinfo {author} {\bibfnamefont {M.~S.}\ \bibnamefont {Dixit}},
  \bibinfo {author} {\bibfnamefont {R.}~\bibnamefont {Dubois}}, \bibinfo
  {author} {\bibfnamefont {J.~A.}\ \bibnamefont {Macdonald}}, \bibinfo {author}
  {\bibfnamefont {T.}~\bibnamefont {Numao}}, \bibinfo {author} {\bibfnamefont
  {B.}~\bibnamefont {Olaniyi}}, \bibinfo {author} {\bibfnamefont
  {A.}~\bibnamefont {Olin}}, \ and\ \bibinfo {author} {\bibfnamefont {J.~M.}\
  \bibnamefont {Poutissou}},\ }\href {\doibase 10.1103/PhysRevD.33.1211}
  {\bibfield  {journal} {\bibinfo  {journal} {Phys. Rev. D}\ }\textbf {\bibinfo
  {volume} {33}},\ \bibinfo {pages} {1211} (\bibinfo {year}
  {1986})}\BibitemShut {NoStop}%
\bibitem [{\citenamefont {Czapek}\ \emph {et~al.}(1993)\citenamefont {Czapek}
  \emph {et~al.}}]{Czapek:1993kc}%
  \BibitemOpen
  \bibfield  {author} {\bibinfo {author} {\bibfnamefont {G.}~\bibnamefont
  {Czapek}} \emph {et~al.},\ }\href {\doibase 10.1103/PhysRevLett.70.17}
  {\bibfield  {journal} {\bibinfo  {journal} {Phys. Rev. Lett.}\ }\textbf
  {\bibinfo {volume} {70}},\ \bibinfo {pages} {17} (\bibinfo {year}
  {1993})}\BibitemShut {NoStop}%
\bibitem [{\citenamefont {Britton}\ \emph {et~al.}(1994)\citenamefont {Britton}
  \emph {et~al.}}]{Britton:1993cj}%
  \BibitemOpen
  \bibfield  {author} {\bibinfo {author} {\bibfnamefont {D.~I.}\ \bibnamefont
  {Britton}} \emph {et~al.},\ }\href {\doibase 10.1103/PhysRevD.49.28}
  {\bibfield  {journal} {\bibinfo  {journal} {Phys. Rev. D}\ }\textbf {\bibinfo
  {volume} {49}},\ \bibinfo {pages} {28} (\bibinfo {year} {1994})}\BibitemShut
  {NoStop}%
\bibitem [{\citenamefont {Aguilar-Arevalo}\ \emph {et~al.}(2015)\citenamefont
  {Aguilar-Arevalo} \emph {et~al.}}]{PiENu:2015seu}%
  \BibitemOpen
  \bibfield  {author} {\bibinfo {author} {\bibfnamefont {A.}~\bibnamefont
  {Aguilar-Arevalo}} \emph {et~al.} (\bibinfo {collaboration} {PiENu}),\ }\href
  {\doibase 10.1103/PhysRevLett.115.071801} {\bibfield  {journal} {\bibinfo
  {journal} {Phys. Rev. Lett.}\ }\textbf {\bibinfo {volume} {115}},\ \bibinfo
  {pages} {071801} (\bibinfo {year} {2015})},\ \Eprint
  {http://arxiv.org/abs/1506.05845} {arXiv:1506.05845 [hep-ex]} \BibitemShut
  {NoStop}%
\bibitem [{\citenamefont {Palomares-Ruiz}\ \emph {et~al.}(2005)\citenamefont
  {Palomares-Ruiz}, \citenamefont {Pascoli},\ and\ \citenamefont
  {Schwetz}}]{Palomares-Ruiz:2005zbh}%
  \BibitemOpen
  \bibfield  {author} {\bibinfo {author} {\bibfnamefont {S.}~\bibnamefont
  {Palomares-Ruiz}}, \bibinfo {author} {\bibfnamefont {S.}~\bibnamefont
  {Pascoli}}, \ and\ \bibinfo {author} {\bibfnamefont {T.}~\bibnamefont
  {Schwetz}},\ }\href {\doibase 10.1088/1126-6708/2005/09/048} {\bibfield
  {journal} {\bibinfo  {journal} {JHEP}\ }\textbf {\bibinfo {volume} {09}},\
  \bibinfo {pages} {048} (\bibinfo {year} {2005})},\ \Eprint
  {http://arxiv.org/abs/hep-ph/0505216} {arXiv:hep-ph/0505216} \BibitemShut
  {NoStop}%
\bibitem [{\citenamefont {Moss}\ \emph {et~al.}(2018)\citenamefont {Moss},
  \citenamefont {Moulai}, \citenamefont {Arg\"uelles},\ and\ \citenamefont
  {Conrad}}]{Moss:2017pur}%
  \BibitemOpen
  \bibfield  {author} {\bibinfo {author} {\bibfnamefont {Z.}~\bibnamefont
  {Moss}}, \bibinfo {author} {\bibfnamefont {M.~H.}\ \bibnamefont {Moulai}},
  \bibinfo {author} {\bibfnamefont {C.~A.}\ \bibnamefont {Arg\"uelles}}, \ and\
  \bibinfo {author} {\bibfnamefont {J.~M.}\ \bibnamefont {Conrad}},\ }\href
  {\doibase 10.1103/PhysRevD.97.055017} {\bibfield  {journal} {\bibinfo
  {journal} {Phys. Rev. D}\ }\textbf {\bibinfo {volume} {97}},\ \bibinfo
  {pages} {055017} (\bibinfo {year} {2018})},\ \Eprint
  {http://arxiv.org/abs/1711.05921} {arXiv:1711.05921 [hep-ph]} \BibitemShut
  {NoStop}%
\bibitem [{\citenamefont {Dentler}\ \emph {et~al.}(2020)\citenamefont
  {Dentler}, \citenamefont {Esteban}, \citenamefont {Kopp},\ and\ \citenamefont
  {Machado}}]{Dentler:2019dhz}%
  \BibitemOpen
  \bibfield  {author} {\bibinfo {author} {\bibfnamefont {M.}~\bibnamefont
  {Dentler}}, \bibinfo {author} {\bibfnamefont {I.}~\bibnamefont {Esteban}},
  \bibinfo {author} {\bibfnamefont {J.}~\bibnamefont {Kopp}}, \ and\ \bibinfo
  {author} {\bibfnamefont {P.}~\bibnamefont {Machado}},\ }\href {\doibase
  10.1103/PhysRevD.101.115013} {\bibfield  {journal} {\bibinfo  {journal}
  {Phys. Rev. D}\ }\textbf {\bibinfo {volume} {101}},\ \bibinfo {pages}
  {115013} (\bibinfo {year} {2020})},\ \Eprint
  {http://arxiv.org/abs/1911.01427} {arXiv:1911.01427 [hep-ph]} \BibitemShut
  {NoStop}%
\bibitem [{\citenamefont {de~Gouv\^ea}\ \emph {et~al.}(2020)\citenamefont
  {de~Gouv\^ea}, \citenamefont {Peres}, \citenamefont {Prakash},\ and\
  \citenamefont {Stenico}}]{deGouvea:2019qre}%
  \BibitemOpen
  \bibfield  {author} {\bibinfo {author} {\bibfnamefont {A.}~\bibnamefont
  {de~Gouv\^ea}}, \bibinfo {author} {\bibfnamefont {O.~L.~G.}\ \bibnamefont
  {Peres}}, \bibinfo {author} {\bibfnamefont {S.}~\bibnamefont {Prakash}}, \
  and\ \bibinfo {author} {\bibfnamefont {G.~V.}\ \bibnamefont {Stenico}},\
  }\href {\doibase 10.1007/JHEP07(2020)141} {\bibfield  {journal} {\bibinfo
  {journal} {JHEP}\ }\textbf {\bibinfo {volume} {07}},\ \bibinfo {pages} {141}
  (\bibinfo {year} {2020})},\ \Eprint {http://arxiv.org/abs/1911.01447}
  {arXiv:1911.01447 [hep-ph]} \BibitemShut {NoStop}%
\bibitem [{\citenamefont {Hostert}\ and\ \citenamefont
  {Pospelov}(2021)}]{Hostert:2020oui}%
  \BibitemOpen
  \bibfield  {author} {\bibinfo {author} {\bibfnamefont {M.}~\bibnamefont
  {Hostert}}\ and\ \bibinfo {author} {\bibfnamefont {M.}~\bibnamefont
  {Pospelov}},\ }\href {\doibase 10.1103/PhysRevD.104.055031} {\bibfield
  {journal} {\bibinfo  {journal} {Phys. Rev. D}\ }\textbf {\bibinfo {volume}
  {104}},\ \bibinfo {pages} {055031} (\bibinfo {year} {2021})},\ \Eprint
  {http://arxiv.org/abs/2008.11851} {arXiv:2008.11851 [hep-ph]} \BibitemShut
  {NoStop}%
\bibitem [{\citenamefont {Berryman}\ \emph {et~al.}(2018)\citenamefont
  {Berryman}, \citenamefont {De~Gouv\^ea}, \citenamefont {Kelly},\ and\
  \citenamefont {Zhang}}]{Berryman:2018ogk}%
  \BibitemOpen
  \bibfield  {author} {\bibinfo {author} {\bibfnamefont {J.~M.}\ \bibnamefont
  {Berryman}}, \bibinfo {author} {\bibfnamefont {A.}~\bibnamefont
  {De~Gouv\^ea}}, \bibinfo {author} {\bibfnamefont {K.~J.}\ \bibnamefont
  {Kelly}}, \ and\ \bibinfo {author} {\bibfnamefont {Y.}~\bibnamefont
  {Zhang}},\ }\href {\doibase 10.1103/PhysRevD.97.075030} {\bibfield  {journal}
  {\bibinfo  {journal} {Phys. Rev. D}\ }\textbf {\bibinfo {volume} {97}},\
  \bibinfo {pages} {075030} (\bibinfo {year} {2018})},\ \Eprint
  {http://arxiv.org/abs/1802.00009} {arXiv:1802.00009 [hep-ph]} \BibitemShut
  {NoStop}%
\bibitem [{\citenamefont {Abbasi}\ \emph {et~al.}(2022)\citenamefont {Abbasi}
  \emph {et~al.}}]{IceCubeCollaboration:2022tso}%
  \BibitemOpen
  \bibfield  {author} {\bibinfo {author} {\bibfnamefont {R.}~\bibnamefont
  {Abbasi}} \emph {et~al.} (\bibinfo {collaboration} {(IceCube Collaboration)*,
  IceCube}),\ }\href {\doibase 10.1103/PhysRevLett.129.151801} {\bibfield
  {journal} {\bibinfo  {journal} {Phys. Rev. Lett.}\ }\textbf {\bibinfo
  {volume} {129}},\ \bibinfo {pages} {151801} (\bibinfo {year} {2022})},\
  \Eprint {http://arxiv.org/abs/2204.00612} {arXiv:2204.00612 [hep-ex]}
  \BibitemShut {NoStop}%
\bibitem [{\citenamefont {Hardin}\ \emph {et~al.}(2022)\citenamefont {Hardin},
  \citenamefont {Martinez-Soler}, \citenamefont {Diaz}, \citenamefont {Jin},
  \citenamefont {Kamp}, \citenamefont {Arg\"uelles}, \citenamefont {Conrad},\
  and\ \citenamefont {Shaevitz}}]{Hardin:2022muu}%
  \BibitemOpen
  \bibfield  {author} {\bibinfo {author} {\bibfnamefont {J.~M.}\ \bibnamefont
  {Hardin}}, \bibinfo {author} {\bibfnamefont {I.}~\bibnamefont
  {Martinez-Soler}}, \bibinfo {author} {\bibfnamefont {A.}~\bibnamefont
  {Diaz}}, \bibinfo {author} {\bibfnamefont {M.}~\bibnamefont {Jin}}, \bibinfo
  {author} {\bibfnamefont {M.~W.}\ \bibnamefont {Kamp}}, \bibinfo {author}
  {\bibfnamefont {C.~A.}\ \bibnamefont {Arg\"uelles}}, \bibinfo {author}
  {\bibfnamefont {J.~M.}\ \bibnamefont {Conrad}}, \ and\ \bibinfo {author}
  {\bibfnamefont {M.~H.}\ \bibnamefont {Shaevitz}},\ }\href@noop {} {\
  (\bibinfo {year} {2022})},\ \Eprint {http://arxiv.org/abs/2211.02610}
  {arXiv:2211.02610 [hep-ph]} \BibitemShut {NoStop}%
\bibitem [{\citenamefont {Antonello}\ \emph {et~al.}(2015)\citenamefont
  {Antonello} \emph {et~al.}}]{MicroBooNE:2015bmn}%
  \BibitemOpen
  \bibfield  {author} {\bibinfo {author} {\bibfnamefont {M.}~\bibnamefont
  {Antonello}} \emph {et~al.} (\bibinfo {collaboration} {MicroBooNE, LAr1-ND,
  ICARUS-WA104}),\ }\href@noop {} {\  (\bibinfo {year} {2015})},\ \Eprint
  {http://arxiv.org/abs/1503.01520} {arXiv:1503.01520 [physics.ins-det]}
  \BibitemShut {NoStop}%
\bibitem [{\citenamefont {Anghel}\ \emph {et~al.}(2015)\citenamefont {Anghel}
  \emph {et~al.}}]{ANNIE:2015inw}%
  \BibitemOpen
  \bibfield  {author} {\bibinfo {author} {\bibfnamefont {I.}~\bibnamefont
  {Anghel}} \emph {et~al.} (\bibinfo {collaboration} {ANNIE}),\ }\href@noop {}
  {\  (\bibinfo {year} {2015})},\ \Eprint {http://arxiv.org/abs/1504.01480}
  {arXiv:1504.01480 [physics.ins-det]} \BibitemShut {NoStop}%
\bibitem [{\citenamefont {Back}\ \emph {et~al.}(2017)\citenamefont {Back} \emph
  {et~al.}}]{ANNIE:2017nng}%
  \BibitemOpen
  \bibfield  {author} {\bibinfo {author} {\bibfnamefont {A.~R.}\ \bibnamefont
  {Back}} \emph {et~al.} (\bibinfo {collaboration} {ANNIE}),\ }\href@noop {} {\
   (\bibinfo {year} {2017})},\ \Eprint {http://arxiv.org/abs/1707.08222}
  {arXiv:1707.08222 [physics.ins-det]} \BibitemShut {NoStop}%
\bibitem [{\citenamefont {Aguilar-Arevalo}\ \emph
  {et~al.}(2021{\natexlab{b}})\citenamefont {Aguilar-Arevalo} \emph
  {et~al.}}]{MiniBooNE:2021bgc}%
  \BibitemOpen
  \bibfield  {author} {\bibinfo {author} {\bibfnamefont {A.~A.}\ \bibnamefont
  {Aguilar-Arevalo}} \emph {et~al.} (\bibinfo {collaboration} {MiniBooNE}),\
  }\href@noop {} {\  (\bibinfo {year} {2021}{\natexlab{b}})},\ \Eprint
  {http://arxiv.org/abs/2110.15055} {arXiv:2110.15055 [hep-ex]} \BibitemShut
  {NoStop}%
\end{thebibliography}%

\appendix

\onecolumngrid

\section{Further Details on the Analysis} \label{app:details}
This supplement to the main text is intended to provide additional detail on the analysis, specifically concerning the calculation of the response matrices used to unfold the MiniBooNE excess and obtain a $\overline{\nu}_e$ prediction in MicroBooNE.

\subsection{Response Matrix Calculation} \label{app:response_matrix}

To approximate $R_{i\alpha}$ in MiniBooNE, we first calculate the conditional probability density function (PDF) $P(E_{\bar{\nu}}^{\rm reco}|E_{\bar{\nu}}^{\rm true})$.
We marginalize over $E_\nu^{\rm QE,true}$, which, due to nuclear effects and differences between the $E_\nu^{\rm QE}$  and $E_{\bar{\nu}}^{\rm QE}$ expressions, is not necessarily the same as the generated antineutrino energy $E_{\bar{\nu}}^{\rm true}$.
We use \textsc{GENIE\,v3.02.00} to discretely approximate $P(E_\nu^{\rm QE,true} | E_{\bar{\nu}}^{\rm true})$ and use a Gaussian approximation for $P(E_\nu^{\rm QE,reco} | E_\nu^{\rm QE,true})$.
Note that since we are considering MiniBooNE neutrino mode data, $E_{\bar{\nu}}^{\rm reco} \equiv E_\nu^{\rm QE,reco}$.
The full calculation of the conditional PDF is
\begin{equation}{\label{eq:conditional_pdf}}
\begin{split}
&P(E_{\bar{\nu}}^{\rm reco} | E_{\bar{\nu}}^{\rm true})
= \int_0^\infty d E_\nu^{\rm QE,true} P(E_\nu^{\rm QE,reco} | E_\nu^{\rm QE,true}) P(E_\nu^{\rm QE,true} | E_{\bar{\nu}}^{\rm true}) \\
&= \sum_{E_\nu^{\rm QE,true}{\rm~bins~}k}
P^{\rm GENIE}_{k} (E_{\bar{\nu}}^{\rm true})
\int_{(E_\nu^{\rm QE,true})_k^{\rm low}}^{(E_\nu^{\rm QE,true})_k^{\rm high}}
d E_\nu^{\rm QE,true} \frac{ \exp \Big[ \frac{-\big(E_\nu^{\rm QE,reco} - E_\nu^{\rm QE,true} \big)^2}{ 2(\sigma(E_{\bar{\nu}}^{\rm true}))^2}  \Big] }{\sqrt{2\pi (\sigma (E_{\bar{\nu}}^{\rm true}))^2}} \\
&= \sum_{E_\nu^{\rm QE,true}{\rm~bins~}k} P^{\rm GENIE}_{k} (E_{\bar{\nu}}^{\rm true}) \frac{1}{2} \Big[ {\rm Erf}\Big( \frac{(E_\nu^{\rm QE,true})_k^{\rm high} - E_\nu^{\rm QE,reco}}{\sqrt{2} \sigma (E_{\bar{\nu}}^{\rm true})} \Big) - {\rm Erf}\Big( \frac{(E_\nu^{\rm QE,true})_k^{\rm low} - E_\nu^{\rm QE,reco}}{\sqrt{2}\sigma (E_{\bar{\nu}}^{\rm true})} \Big) \Big],
\end{split}
\end{equation}
where ${\rm Erf}$ denotes the error function and we define $P^{\rm GENIE}_{k}(E_{\bar{\nu}}^{\rm true}) \equiv P((E_\nu^{\rm QE,true})_k | E_{\bar{\nu}}^{\rm true})$ to be the binned $E_\nu^{\rm QE,true}$ probability distribution calculated via \textsc{GENIE\,v3.02.00} for a given $E_{\bar{\nu}}^{\rm true}$.
In the third line, we convert the continuous integral over $E_\nu^{\rm QE,true}$ to a discrete sum over the binned $P^{\rm GENIE}_{k\alpha}$ distribution and explicitly show our Gaussian approximation, where we consider a flat 17\% energy resolution, i.e., $\sigma(E_{\bar{\nu}}^{\rm true}) = 0.17 E_{\bar{\nu}}^{\rm true}$.
We also consider a flat 5\% $E_\nu^{\rm QE}$ underestimation bias.
This energy resolution and bias are derived from the most recent MiniBooNE $\nu_e$ data release~\cite{MiniBooNE:2021bgc}; for more details see \cref{app:approximations}.
In the fourth line, we integrate the Gaussian within each $E_\nu^{\rm QE,true}$ bin $k$.
We can then approximate $R_{i\alpha}$ by integrating $P(E_{\bar{\nu}}^{\rm reco} | E_{\bar{\nu}}^{\rm true})$ over $E_{\bar{\nu}}^{\rm reco}$ bin $i$ and averaging over $E_{\bar{\nu}}^{\rm true}$ bin $\alpha$,
\begin{equation}{\label{eq:response_matrix}}
R_{i\alpha} = \frac{1}{(E_{\bar{\nu}}^{\rm true})_\alpha^{\rm high} - (E_{\bar{\nu}}^{\rm true})_\alpha^{\rm low}}
\int_{(E_{\bar{\nu}}^{\rm true})_\alpha^{\rm low}}^{(E_{\bar{\nu}}^{\rm true})_\alpha^{\rm high}}
\int_{(E_{\bar{\nu}}^{\rm reco})_i^{\rm low}}^{(E_{\bar{\nu}}^{\rm reco})_i^{\rm high}}
d E_{\bar{\nu}}^{\rm reco} d E_{\bar{\nu}}^{\rm true} P(E_{\bar{\nu}}^{\rm reco} | E_{\bar{\nu}}^{\rm true}).
\end{equation}

\begin{figure}[t]
    \centering
    \includegraphics[width=0.45\textwidth]{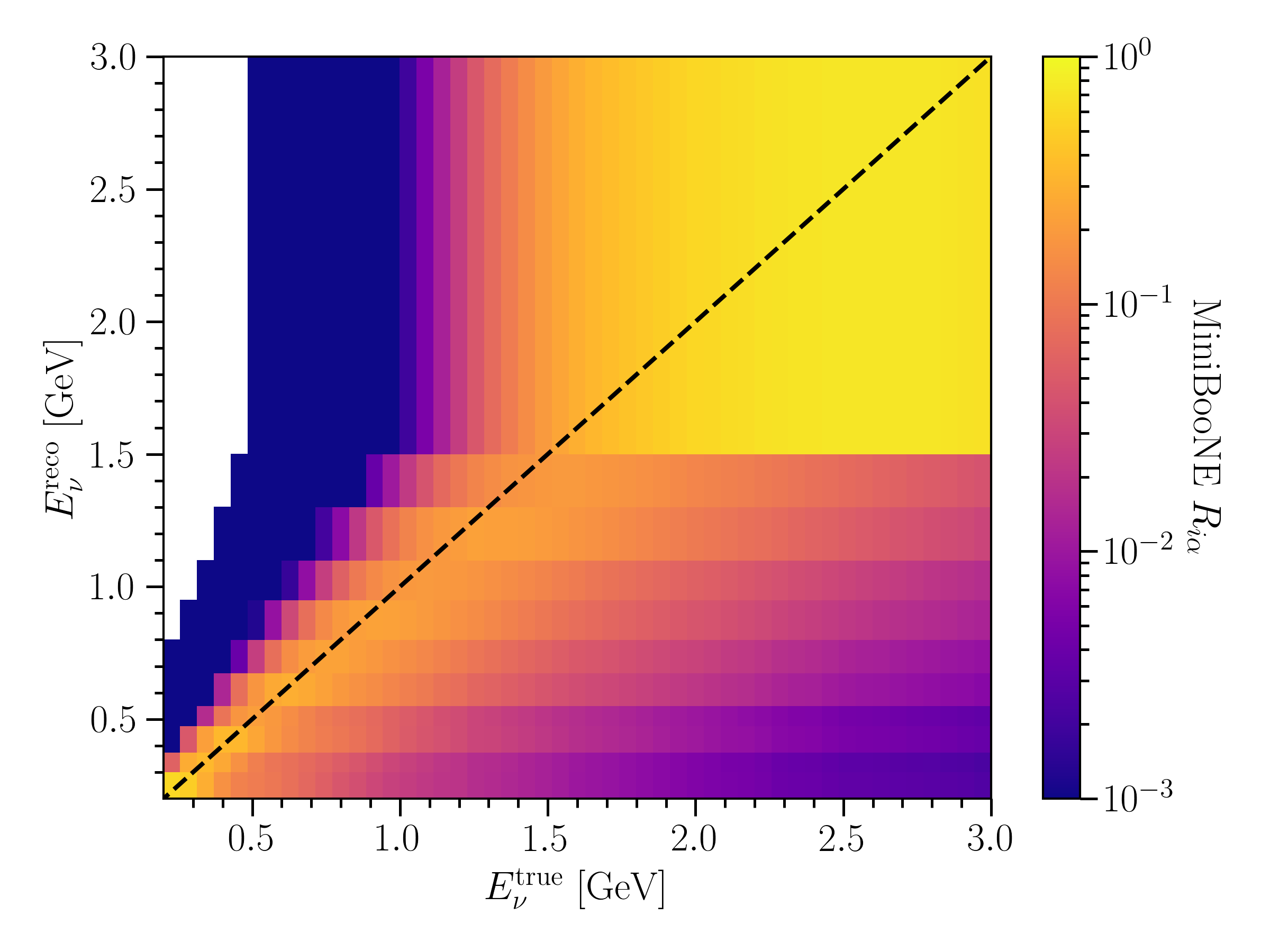}
    \includegraphics[width=0.45\textwidth]{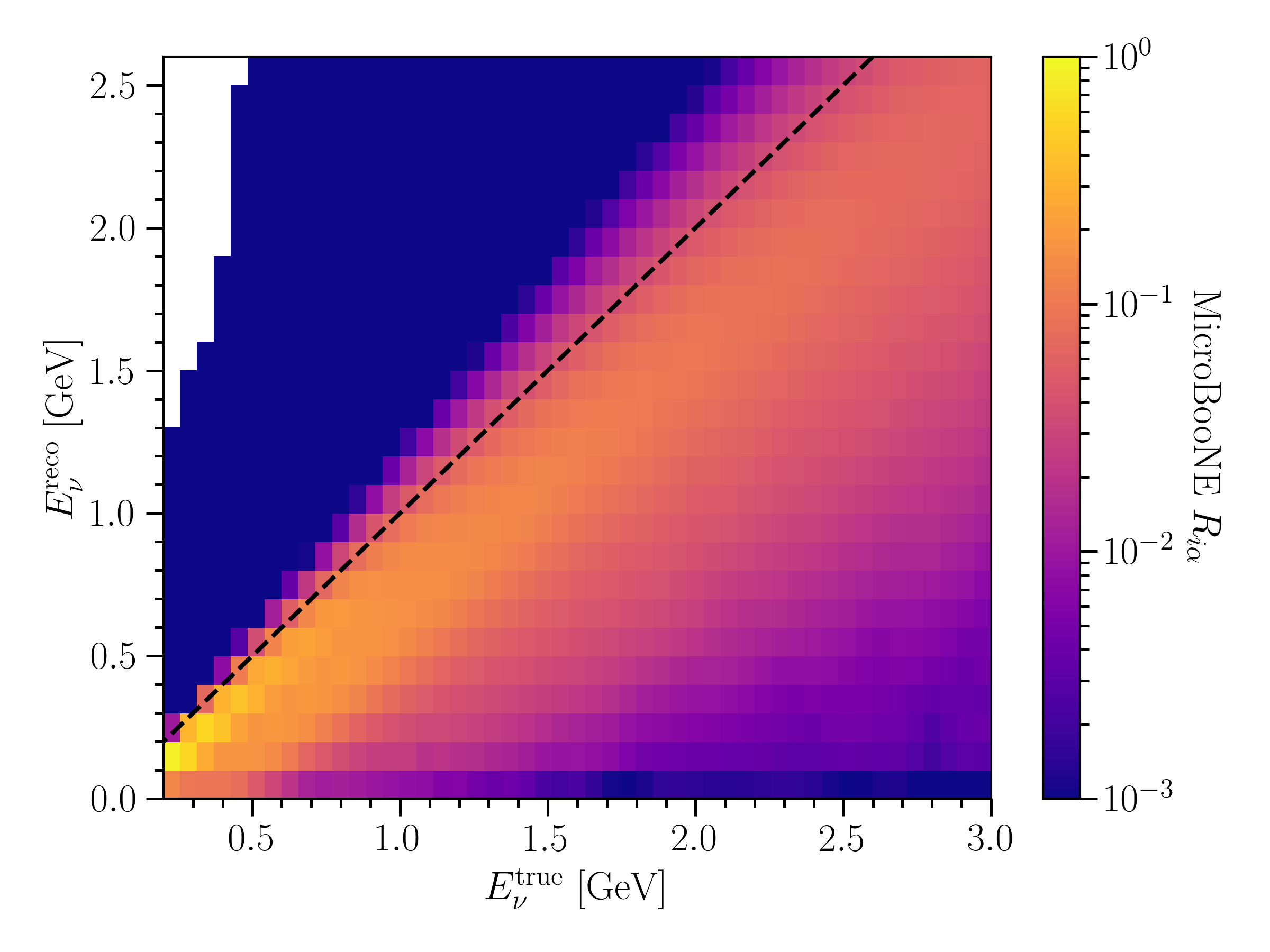}
    \caption{MiniBooNE (top) and MicroBooNE (bottom) $\overline{\nu}_e$ response matrices $R_{i\alpha}$, calculated according to \cref{app:response_matrix}. Reconstructed energy binning in each matrix reflects the binning reported by each collaboration.}
    \label{fig:response_matrices}
\end{figure}

The $\overline{\nu}_e$ response matrix in MicroBooNE begins with a calculation similar to \cref{eq:conditional_pdf}.
Due to the calorimetric energy reconstruction used in the WireCell analysis, the calculation of the conditional PDF $P(E_{\overline{\nu}}^{\rm reco} | E_{\overline{\nu}}^{\rm true})$ in MicroBooNE marginalizes over the true positron kinetic energy,
\begin{equation}{\label{eq:microboone_conditional_pdf}}
P(E_{\overline{\nu}}^{\rm reco} | E_{\overline{\nu}}^{\rm true})
= \int_0^\infty d T_{e^+}^{\rm true} P(E_{\overline{\nu}}^{\rm Cal,reco} | T_{e^+}^{\rm true}) P(T_{e^+}^{\rm true} | E_{\overline{\nu}}^{\rm true}).
\end{equation}
We perform the same procedure used for MiniBooNE to approximate $R_{i \alpha}$ in MicroBooNE, leveraging \textsc{GENIE\,v3.02.00} to calculate $P(T_{e^+}^{\rm true} | E_{\overline{\nu}}^{\rm true})$ and making a Gaussian approximation for $P(E_{\overline{\nu}}^{\rm Cal,reco} | T_{e^+}^{\rm true})$.
Following Ref.~\cite{MicroBooNE:2021ojx}, we consider a flat 2\% relative bias and 12\% relative uncertainty on the lepton kinetic energy.
The MiniBooNE and MicroBooNE $\bar{\nu}_e$ response matrices calculated using \cref{eq:conditional_pdf,eq:microboone_conditional_pdf,eq:response_matrix} are shown in \cref{fig:response_matrices}.
The significant population of events below the $E_{\overline{\nu}}^{\rm reco} = E_{\overline{\nu}}^{\rm true}$ line in the MicroBooNE matrix indicates the under-estimation bias from the invisible neutron.

\subsection{Approximations in Response Matrices} \label{app:approximations}

\begin{figure}[th]
    \centering
    \includegraphics[width=0.45\textwidth]{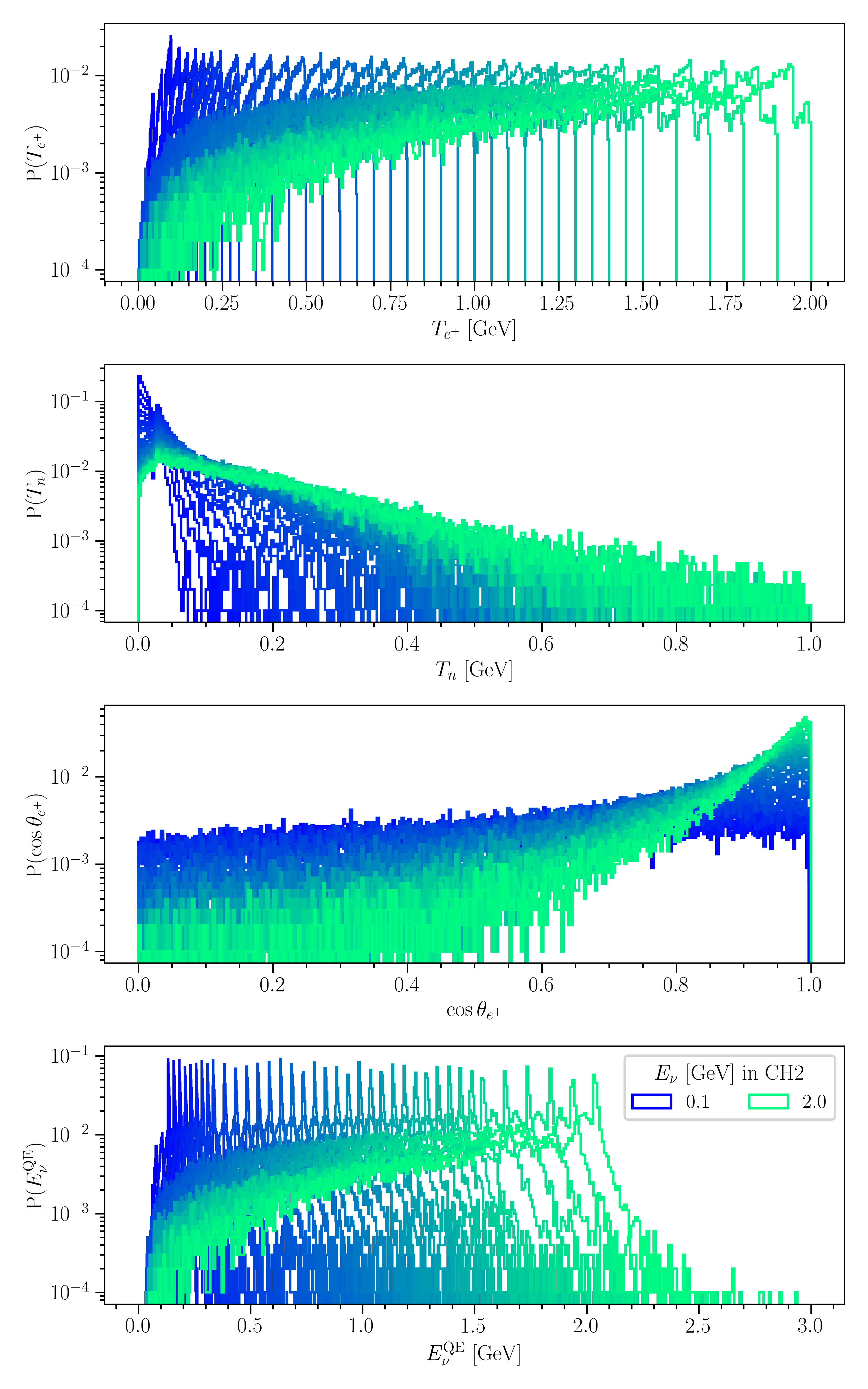}
    \includegraphics[width=0.45\textwidth]{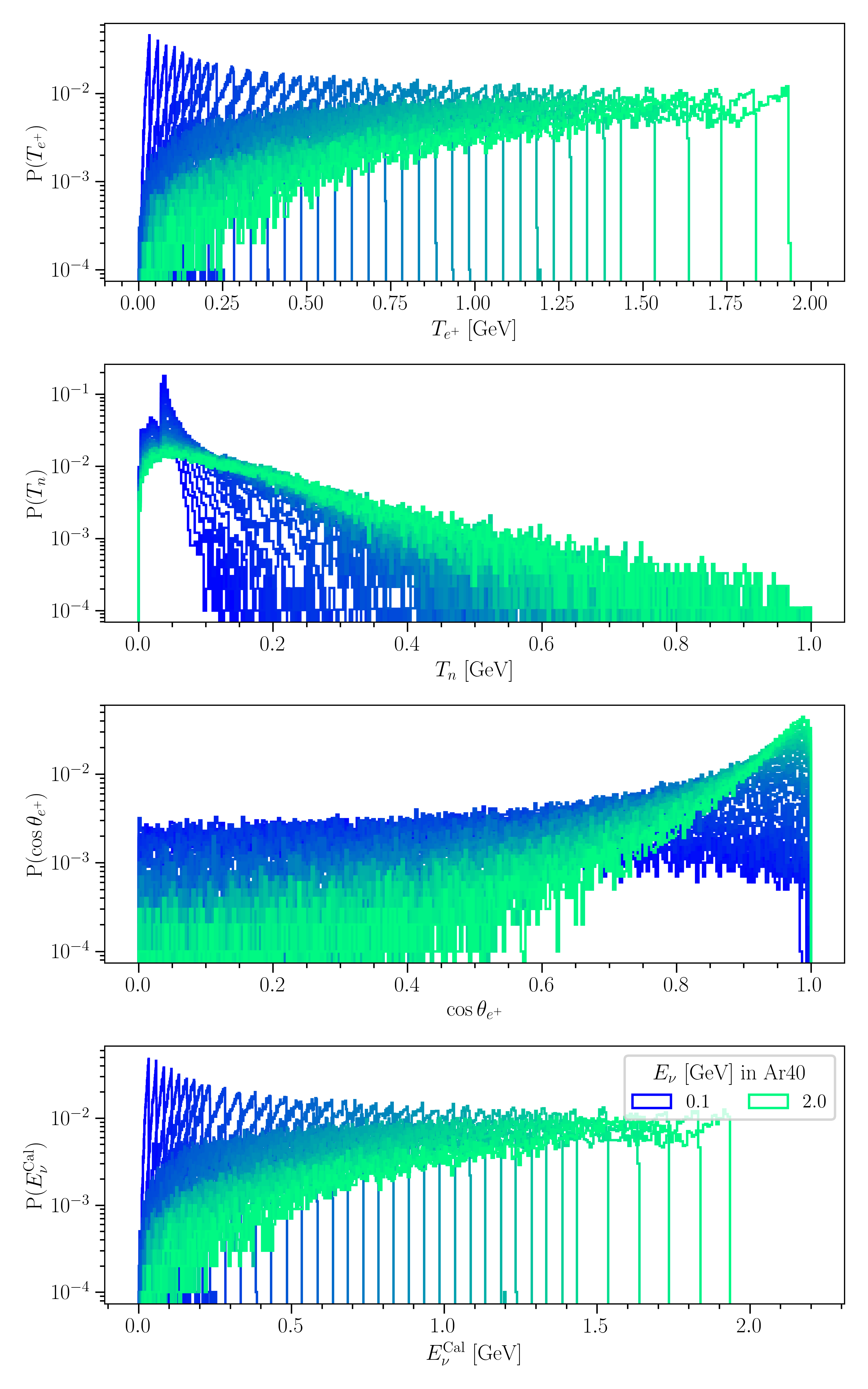}
    \caption{Truth-level distributions of final state kinematic variables for $\bar{\nu}_e$ scattering in CH$_2$ (left) and Ar40 (right), generated using \textsc{GENIE\,v3.02.00}.}
    \label{fig:GENIE}
\end{figure}

To approximate the truth-level PDFs in \cref{eq:conditional_pdf} and \cref{eq:microboone_conditional_pdf}, we use the latest version of the GENIE event generator, \textsc{GENIE\,v3.02.00}~\cite{Andreopoulos:2015wxa}.
As the hypotheses presented in this study attribute at least part of the MiniBooNE excess to CC~$\bar{\nu}_e$ scattering, we generate all charged-current $\bar{\nu}_e$ interactions included in \textsc{GENIE\,v3.02.00} in both CH$_2$ and Ar40.
Truth-level distributions of final state variables in these interactions are shown in \cref{fig:GENIE}, including the positron kinetic energy $T_{e^+}$, neutron kinetic energy $T_n$, and positron scattering angle $\cos \theta_{e^+}$.
We also show the truth-level distribution of the energy reconstruction definition for each experiment--$E_\nu^{\rm QE}$ for MiniBooNE and $E_\nu^{\rm Cal}$ for MicroBooNE.
Note that in the MicroBooNE case, the $E_\nu^{\rm Cal}$ distribution matches the $T_{e^+}$ distribution, as the final state neutron is invisible.
In principle, final state protons and charged pions generated in $\bar{\nu}_e$ scattering via interactions within nuclear medium will also contribute to $E_\nu^{\rm Cal}$; however, we have ignored this effect for the purposes of this study.

\begin{figure}[th]
    \centering
    \includegraphics[width=0.45\textwidth]{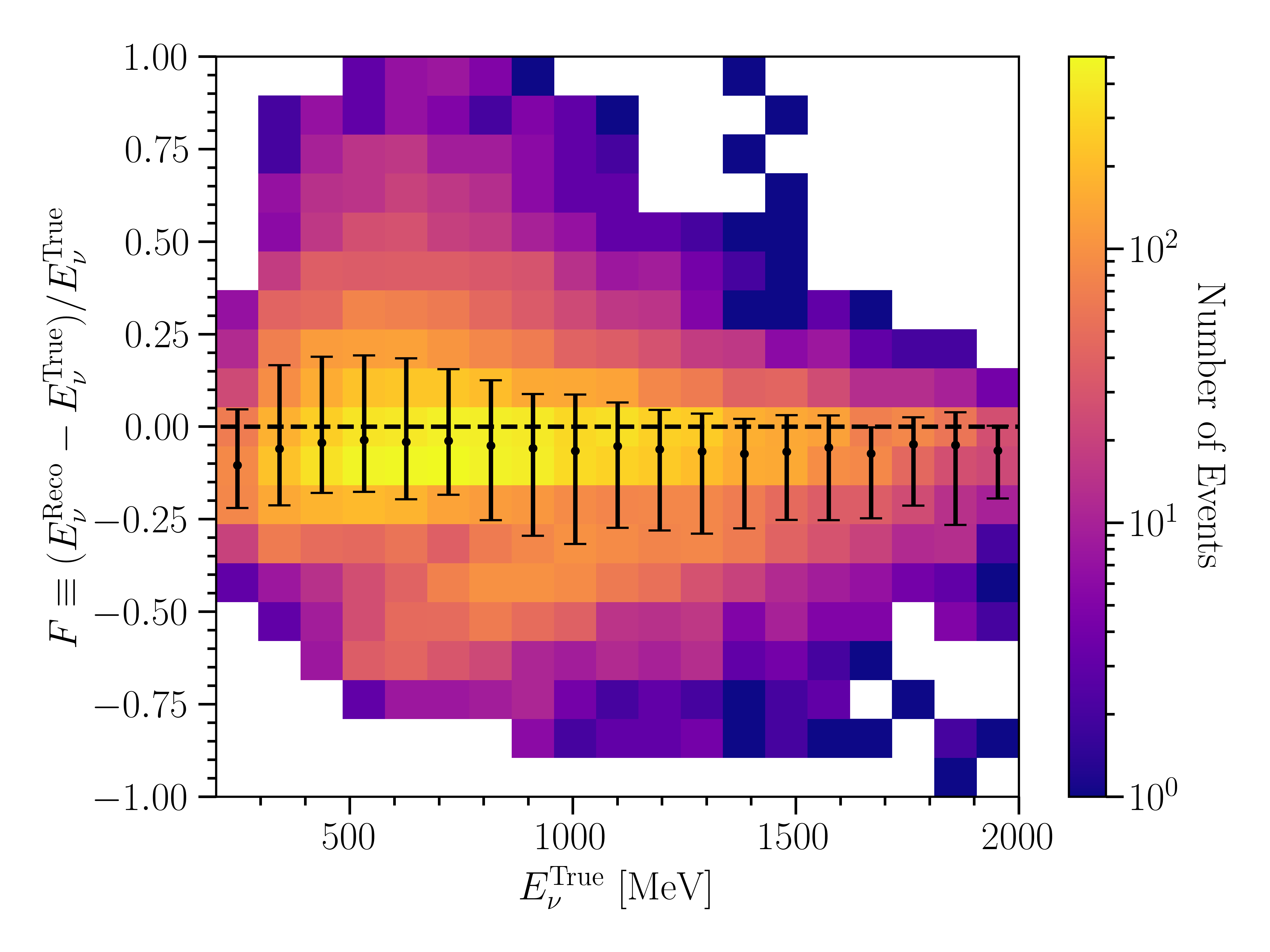}
    \includegraphics[width=0.45\textwidth]{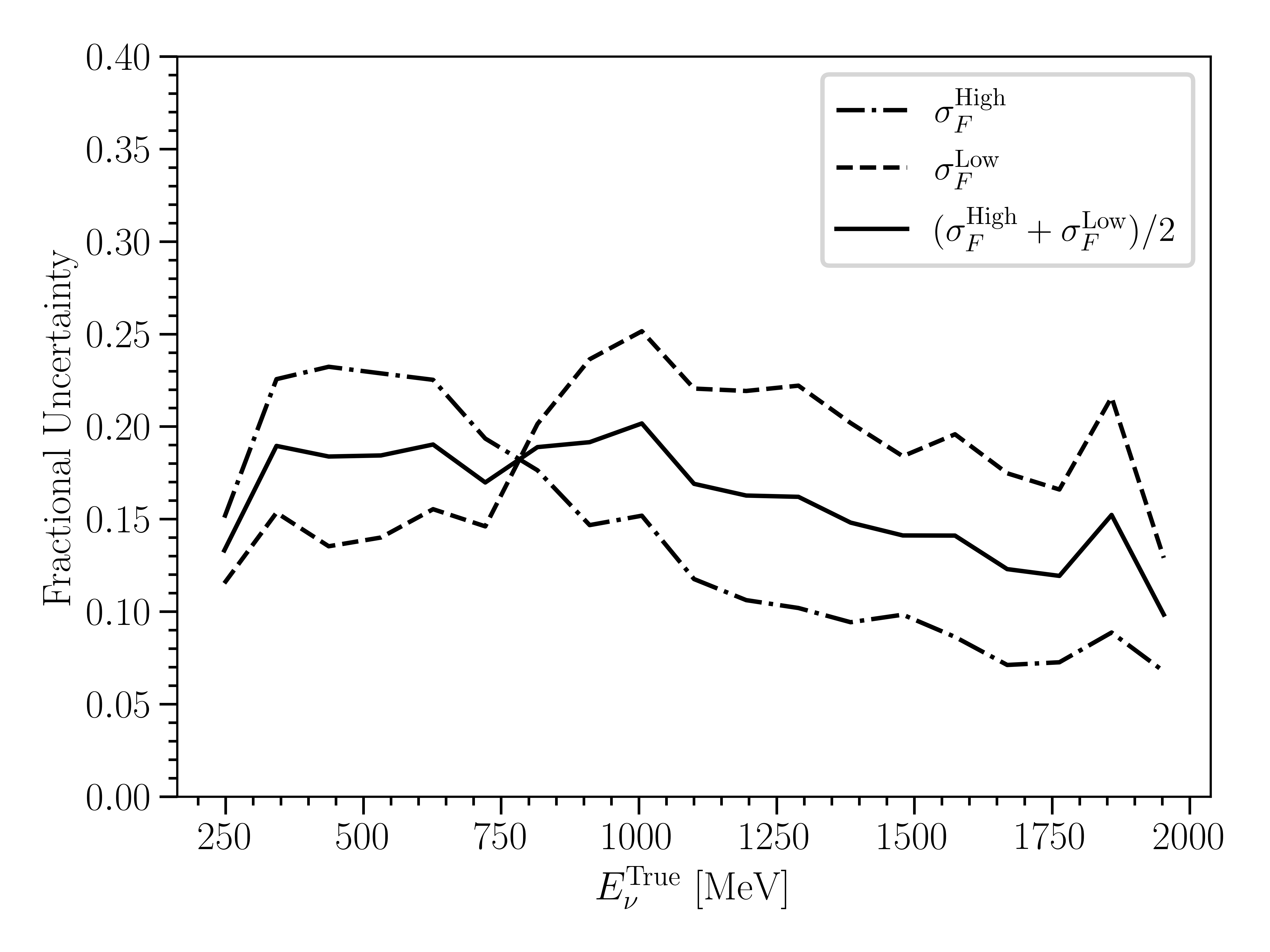}
    \caption{Left: 2D distribution of the fractional energy reconstruction error $F$ as a function of the true neutrino energy. The black points and error bars indicate the median and $\pm 1\sigma$ extent, respectively, of the $F$ distribution in each $E_\nu^{\rm True}$ bin. Right: upper and lower uncertainties on $F$ as a function of the true neutrino energy. The average of the two is also shown.}
    \label{fig:MBEres}
\end{figure}

The MiniBooNE energy resolution is approximated using the latest $\nu_\mu \to \nu_e$ Monte Carlo data release from the MiniBooNE collaboration~\cite{MiniBooNE:2021bgc}.
In \cref{fig:MBEres}, we show a 2D histogram of the fractional error of each event,
\begin{equation}
F \equiv \frac{E_\nu^{\rm Reco} - E_\nu^{\rm True}}{E_\nu^{\rm True}},
\end{equation}
as a function of the true neutrino energy.
A profiled version of this distribution is overlaid in black, where the data points and error bars indicate the median and $\pm 1\sigma$ extent, respectively, of the $F$ distribution in each $E_\nu^{\rm True}$ bin.
From the median, one can see that MiniBooNE tends to consistently under-predict the $\nu_e$ energy.
The fractional uncertainty on the neutrino energy is also relatively constant across the relevant energy range, though it is not symmetric.
This is shown in the right panel of \cref{fig:MBEres}, which plots the upper and lower $1\sigma$ fractional uncertainty on the neutrino energy as a function of $E_\nu^{\rm True}$, as well as the average of the two.
It is apparent that the uncertainty on $F$ is relatively flat across the relevant true neutrino energy range.
One can see that for $E_\nu^{\rm True} < 750\;{\rm MeV}$, the $F$ distribution has a larger extend above the median, while the opposite is true for $E_\nu^{\rm True} > 750\;{\rm MeV}$.
While this is an interesting effect, for the purpose of this study we approximate the MiniBooNE $E_\nu$ fractional uncertainty to be 17\%--the average value $\sigma_F$ across the full neutrino energy range.
We also incorporate a 5\% under-prediction bias, which is the median value of $F$ across the full neutrino energy range.
In principle, $\bar{\nu}_e$ scattering will behave differently than $\nu_e$ scattering in MiniBooNE, as the momentum transfer $Q^2$ distribution and final state lepton kinematics differ between the two.
This effect is accounted for in the construction of the response matrices for both MiniBooNE and MicroBooNE--however, we ignore it when approximating the MiniBooNE energy resolution, as it is a subdominant effect here.
The fractional uncertainty on positron EM showers in MicroBooNE is taken to be 12\% with a 2\% under-prediction bias, as quoted in Ref.~\cite{MicroBooNE:2021ojx}.

\begin{figure}[h]
    \centering
    \includegraphics[width=0.45\textwidth]{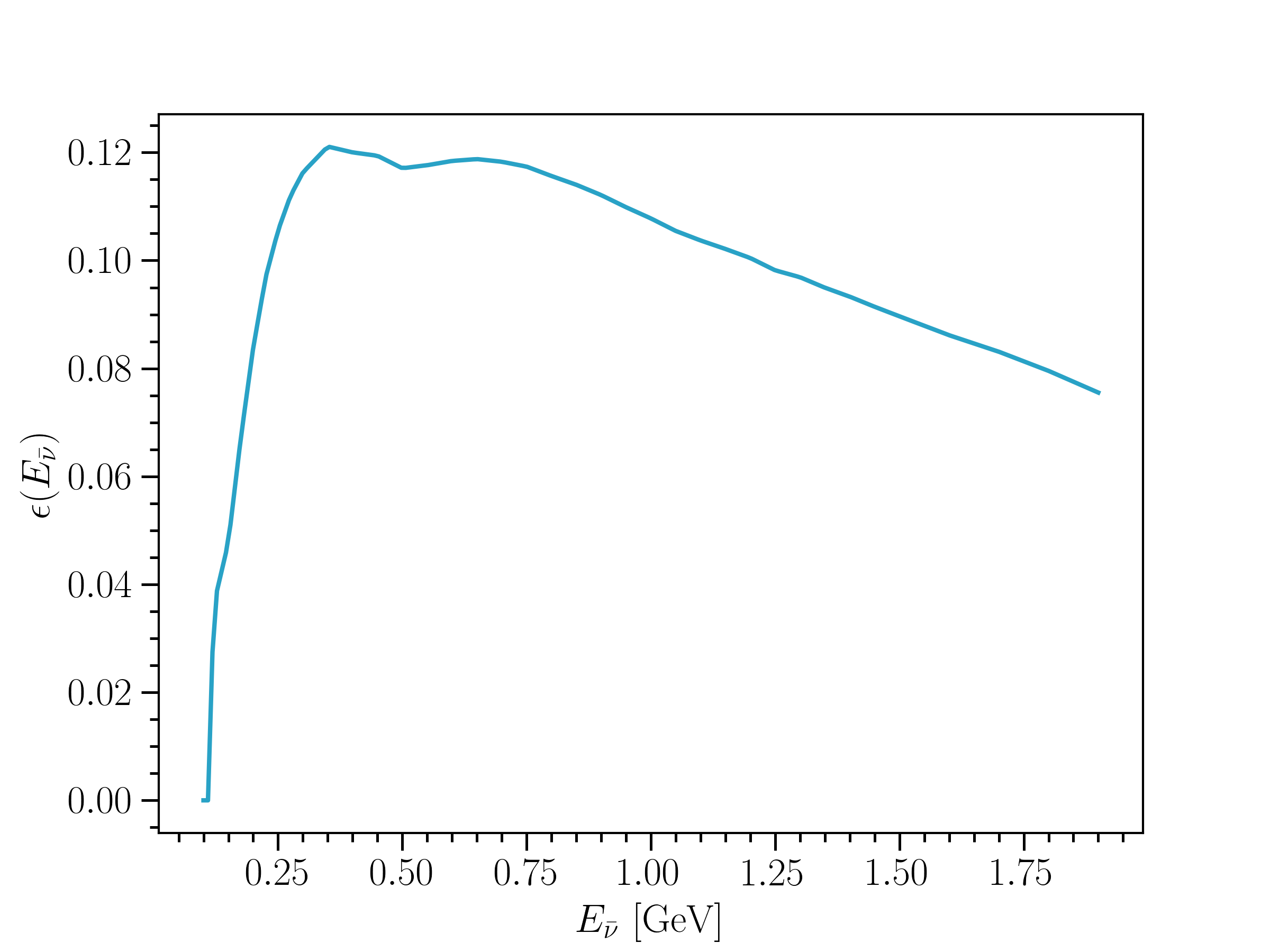}
    \caption{MiniBooNE $\bar{\nu}_e$ detection efficiency as a function of the true antineutrino energy.}
    \label{fig:MB_eff}
\end{figure}

As mentioned in the main text, we use the provided electron reconstruction efficiency $\epsilon(E_{e^+})$ in MiniBooNE~\cite{MiniBooNE_eff} to approximate the $\overline{\nu}_e$ reconstruction efficiency $\epsilon_\alpha$.
As MiniBooNE is a spherically symmetric detector that can only reconstruct the final state lepton, the $e^+$ energy is the dominant effect in the (anti)neutrino detection efficiency.
Using \textsc{GENIE\,v3.02.00}, we can estimate $P(E_{e^+}|E_{\overline{\nu}})$ as shown in \cref{fig:GENIE} and thus approximate the lepton energy-averaged detection efficiency,
\begin{equation}{\label{eq:MB_eff}}
\epsilon(E_{\overline{\nu}}) = \int_0^\infty \epsilon(E_{e^+}) P(E_{e^+}|E_{\overline{\nu}}) dE_{e^+}.
\end{equation}
This efficiency is shown in \cref{fig:MB_eff}.
For reasons discussed in the main text, we consider the $\bar{\nu}_e$ detection efficiency in MicroBooNE to be the same as the $\nu_e$ detection efficiency released by the collaboration.
This likely overestimates the $\bar{\nu}_e$ rate in MicroBooNE, as the neutron created in $\bar{\nu}_e$ CC interactions is not visible in the detector.
This would make our estimation of MicroBooNE's sensitivity to a $\bar{\nu}_e$-generated excess artificially strong
Thus, it is a conservative assumption, given the conclusion of this paper--that MicroBooNE is not sensitive to a $\bar{\nu}_e$-generated excess.

\end{document}